\documentclass[useAMS,usenatbib]{mn2e}

\usepackage{lscape,graphicx,pifont}
\usepackage{times}

\usepackage{longtable}
\usepackage{natbib}
\usepackage{ftnxtra}
\usepackage{fnpos}

\usepackage{color}

\DeclareGraphicsExtensions{.png,.pdf,.eps,.jpg,.tiff}

\newcommand{\Msun}{\mbox{\,M$_\odot$}}

\newcommand{\vunit}{\mbox{\,km\,s$^{-1}$}}
\newcommand{\mic}{\mbox{$\,\mu$m}}
\newcommand{\pion}[2]{{#1}\,{\sc {#2}}}

\newcommand{\ltsimeq}{\raisebox{-0.6ex}{$\,\stackrel
        {\raisebox{-.2ex}{$\textstyle <$}}{\sim}\,$}}
\newcommand{\gtsimeq}{\raisebox{-0.6ex}{$\,\stackrel
        {\raisebox{-.2ex}{$\textstyle >$}}{\sim}\,$}}

\newcommand{\spitzer}{\mbox{\it Spitzer Space Telescope}}

\newcommand{\ddotsec}{\raisebox{-0.6ex}{$\stackrel
  {\raisebox{-.2ex}{s}}{.}$}}
\newcommand{\ddotarcsec}{\mbox{$''\!\!.$}}

\title[CO in SN 2016adj]{Early formation of carbon monoxide in the Centaurus A supernova SN~2016adj}
\author[D. P. K. Banerjee et al.]{D. P. K. Banerjee$^{1}$\thanks{E-mail: orion@prl.res.in},
Vishal Joshi$^1$, 
A. Evans$^2$, 
Mudit Srivastava$^1$, 
N. M. Ashok$^1$, \newauthor 
R. D. Gehrz$^3$, 
M. S. Connelley$^4$, 
T. R. Geballe$^5$, 
J. Spyromilio$^6$, 
J. Rho$^{7,8}$,
R. Roy$^9$\thanks{Present address: Inter-University Centre for Astronomy and Astrophysics, Pune, India} \\
{$^1$}{Physical Research Laboratory, Navrangpura,  Ahmedabad, Gujarat 380009, India}\\
{$^2$}Astrophysics Group, Lennard Jones Laboratory, Keele University, Keele, Staffordshire, ST5 5BG, UK\\
{$^3$}Minnesota Institute for Astrophysics, School of Physics \& Astronomy,
116 Church Street SE, University of Minnesota, Minneapolis, MN 55455, USA\\
{$^4$}{Institute for Astronomy, 640 North A'ohoku Place, Hilo, HI 96720 USA}\\
{$^5$}{Gemini Observatory, 670 N. A'ohoku Place, Hilo, HI, 96720-2700 , USA}\\
{$^6$}{European Southern Observatory, Karl-Schwarzschild-Strasse, 2, D-85748 Garching bei M\"unchen, Germany}\\
{$^7$}{SOFIA Science Center, NASA Ames Research Center, MS211-1, Moffett Field, CA 94043, USA}\\
{$^8$}{SETI Institute, 189 N. Bernardo Ave., Mountain View, CA 94043, USA} \\
{$^9$}{The Oskar Klein Centre, Department of Astronomy, Stockholm University, AlbaNova, 10691 Stockholm, Sweden.}
}

\begin{document}

\date{Version of \today}

\pagerange{\pageref{firstpage}--\pageref{lastpage}} \pubyear{2018}

\maketitle

\label{firstpage}

\begin{abstract}
We present near-infrared spectroscopy of the NGC\,5128 supernova SN~2016adj in the first 2 months following discovery. 
We report the detection of first overtone carbon monoxide emission at $\sim58.2$~d after discovery, 
one of the earliest detections of CO in an erupting supernova. 
We model the CO emission to derive the CO mass, temperature 
and velocity, assuming both pure $^{12}$CO and a composition that includes $^{13}$CO;
{the case for the latter is the
isotopic analyses of meteoritic grains, which suggest that core collapse supernovae can synthesise 
significant amounts of $^{13}$C.}
Our models show that, while the CO data are adequately explained by pure $^{12}$CO, they do not preclude 
the presence of $^{13}$CO, to a limit of $^{12}$C/$^{13}$C$>3$, the first constraint on the 
$^{12}$C/$^{13}$C ratio determined from near-infrared observations.
We estimate the reddening to the object, 
and the effective temperature from the energy distribution at outburst.
We discuss whether the ejecta of SN~2016adj may be carbon-rich, what the infrared
data tell us about the classification of this  supernova,
{and what implications the early formation of CO in supernovae may have for CO
formation in supernovae in general.}
\end{abstract}



\section{Introduction}
\label{intro}

Stars with mass greater than 8\Msun\  end their lives as core-collapse supernovae (CC SNe). 
SNe coming from stars that retain most of their massive hydrogen envelopes until they explode are classified as Type II,
whereas SNe resulting from stars that had lost their H envelopes before they collapsed are called stripped envelope CC SNe
\citep[see][]{b5,b13a}. The loss of the envelope is believed to be due either to stellar winds, or to interaction with a 
companion star \citep[e.g.][]{smith}. CC SNe progenitors with only small amounts of hydrogen remaining in their outer envelopes 
explode as SNe IIb, a classification so-named because their spectra undergo a transition from Type II at early times to 
Type Ib at late times. There is a further sub-division depending on whether 
the He envelope was still present (Type Ib) or if it had been lost (Type Ic).

{The large quantities of dust observed in high-redshift galaxies raises the fundamental 
astrophysical question of the origin of 
dust in the Universe \citep[e.g.][]{isaak02,bertoldi03,gall}.
Whether SNe are a significant source of dust has been a long-standing debate.} 
In the {\em local} Universe it is known that the bulk of observed dust originates in evolved, 
low-mass stars {\citep*[see e.g.][]{gehrz, sarangi}}. However, this can not be the case for the copious 
amount of dust seen in high-redshift
galaxies, as low-mass stars will not have had time to evolve to the dust-producing stage.
On the other hand the SN explosion of massive stars can occur within a few millions of
years after the onset of star formation, suggesting that 
SNe must be a significant source of dust in the early Universe. 
Support for this view comes from the fact that four young Local Group SN remnants, 
SN~1987A \citep{matsuura11,matsuura15}, the Crab
Nebula \citep{gomez}, Cas~A \citep{delooze} and G54.1+0.3 \citep{temim,rho18a} have dust
masses of $0.1-0.9$\Msun.
These results reinforce the conclusion that SNe are important dust factories in the local
Universe, and must be so at high redshift.

The formation of small molecules such as carbon monoxide (CO) is a necessary prerequisite for dust formation
{and indeed,} {CO is one of the most powerful coolants in the ejecta of Type II SNe. 
The outcomes of dust evolution models depend on CO destruction processes, including impact by
the energetic electrons produced by the radioactive decay of $^{56}$Co, and microscopic mixing 
of He$^+$ \citep{cherchneff10}.}

However, detections of first overtone CO emission in SNe are limited in number, 
while the fundamental band -- for which  the transition rates are 
$\sim100\times$ higher than the first overtone --
has been observed only in SN~1987A from the ground \citep{meikle0,catchpole}, 
from the Kuiper Airborne Observatory \citep{wooden}, and in a small number of objects using the
\spitzer\ \citep[e.g.][]{kotak5,kotak6,meikle06,meikle07,fox}. 
The detection of CO first overtone and fundamental bands in the SN remnant Cas~A
\citep{rho09,rho12}, {which \citeauthor{rho12} conclude was formed in an earlier phase of the SN 
explosion, suggests that} astrochemical processes and molecule formation may continue at 
least $\sim300$~yr after the initial explosion, albeit in different environments.

In this paper we present near-IR (NIR) spectroscopic observations of the CC SN~2016adj.
{The observations are at high temporal cadence and moderate ($R\sim1000$)
spectral resolution, and provide an exceptional record of the early development of the IR spectrum
of an erupting SN. Furthermore, our data enable us to}
report one of the earliest detections of first overtone CO emission in an erupting SN. 
{In the past the Mt Abu group has worked extensively on NIR studies of classical novae, 
a major result of which has been the early detection of CO emission and dust formation in novae. 
Because of the modest size of the telescope (1.2~m),  NIR spectroscopy of SNe is generally 
not possible unless the erupting SN is sufficiently bright in the NIR and likely to remain so as 
CO and dust formation occur.  SN~2016adj was bright in the NIR at outburst \citep*[$J,H,K\sim10$~mag;][]{b37}
and early reports indicated it to be a CCSN, many of  which -- as discussed below --  
progress to show CO emission and dust formation. This made SN~2016adj a potentially interesting
candidate to monitor at Mt Abu. The advantage of an institutional telescope that permits high-cadence observations
also provided the potential to catch an early formation of either CO or dust.}

\section{CO and dust formation in supernovae}
\label{COdust}
CO emission in an erupting supernova was first seen
(at $\sim100$~days after the explosion) in SN~1987A \citep{b2, mcgregor, meikle0, catchpole, oliva, b40}.
The SiO fundamental at $\sim8$\mic\ was also observed, by \cite*{roche}. 

In SN~1987A CO emission was seen to be at temperatures (2000~K) below those of the atomic species,
and at velocities (2000\vunit) that were also lower than the velocities of 
some of the metals \citep[][and references therein]{wliu}. 
These observations placed the CO and SiO deep inside the ejecta,
and imposed constraints on the mixing of the ejecta and the
shielding of the molecules from ultraviolet (UV) radiation that would otherwise destroy them.

Since then first overtone CO emission has been observed in a number of
Type II SNe, for example SN~1995ad \citep{SL}, SN~1998S \citep{b14, b12}, SNe~1998dl and 1999em
\citep{b42} and Type IIb SN~2011dh \citep{b11}. Detections of CO in
Type Ic have been made in  SN~2000ew \citep{b15}, SN~2007gr \citep{b17}
and the Type Ib/c SN~2013ge \citep{drout}.  The earliest
detection to date of first overtone CO emission was in SN~2013ge at +48~d after
$B$(max) (58.9~d after discovery), by \cite{drout}.

CO plays a vital role in dust formation because of its formidable cooling capacity,
by emission in the fundamental band at 4.7\mic\ and first overtone bands at 2.29\mic. 
Molecule formation in the early phase of the eruption is conditional on them being shielded from the UV
radiation field \citep*{lepp} and therefore a degree of clumping in the ejecta is required. 
These clumps cool down faster than the surroundings and are likely sites of dust condensation. 
Models of CO formation predict that the CO abundance should decrease from $\sim600$~days
to $\sim1000$~days \citep{cherchneff1,lazzati} for fully mixed case, and that as CO mass
decreases, carbon dust forms.

After only 500~days, there were three cardinal indications that 
SN~1987A had produced a signifcant amount of dust in its ejecta.
These were the extinction of the red wings of the line profiles by dust obscuration, a
downward transition in the $U\!BV$ light curves -- similar to that seem in the light curves of
dust-forming novae -- and the production of significant thermal infrared (IR) emission from
the dust \citep{GN,lucy,wooden}. As the SN aged and cooled, the molecular emission in the IR vanished. 

About 0.4--0.7\Msun\  of dust was detected at the location of SN~1987A by 
\cite{matsuura11,matsuura15} at an age of 8500 days (23.5 years after the eruption) using 
the {\it Herschel Space Observatory}, and were localised in the ejecta of the
SN using the {\it Atacama Large Milli\-metre Array} (ALMA) by \cite{indeb}. 
\cite{kamenetzky}, using ALMA and
{\it Herschel} data, reported emission from CO and SiO rotational transitions 
arising in material moving at velocities comparable to those detected in the early data.

\begin{figure}
\includegraphics[width=8cm,keepaspectratio]{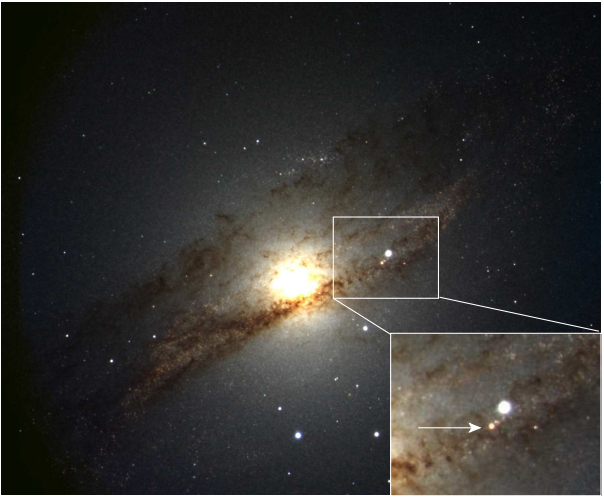}
\caption{A composite $J\!H\!K_{\rm s}$ image of Cen A, showing the location of the SN (arrowed),  
taken on 2016 Aug 7 (+180.4d) with the  Gemini-South telescope. 
The bright star NW of the SN is 2MASS 13252458--4300485. North is up, East is left.
\label{cena}}
\end{figure}

\section{Supernova SN 2016\mbox{adj}}
\label{adj}
SN~2016adj was discovered by \cite{b25} on 2016 February 8.56 UT at $V=14.0$; the SN was not present in an
image taken by them  on 2016 February 3, to a limiting magnitude of $V = 18.0$.
{NIR photometry, obtained within a day of discovery \citep{b37}, revealed a highly 
reddened object.} 

Since all post-discovery optical $B$ and $V$ magnitudes show the SN to fade further
\citep{b19, b39} compared to its discovery magnitude, optical maximum was reached in the period 2016 February 3--8.56.
While it is ideally desirable to consistently measure all elapsed time intervals with respect to (for example)
$B$(max) or $V$(max) -- so that the evolution of SNe may be directly compared --
these are poorly constrained and instead, we use here the date of discovery as the origin of time.

SN~2016adj lies on the central dust lane of the nearby Sy2 galaxy NGC 5128 (Centaurus A; $z= 0.00183$;
SIMBAD\footnote{http://simbad.u-strasbg.fr/simbad/})
in the north-west direction (see Fig.~\ref{cena}).
NGC 5128, an iconic object in itself,  was also  host to the bright Type Ia SN (SN~1986G) 
which occurred in the south-west part of the dust lane.

SN~2016adj has been the subject of considerable interest,  
with reports of special note being those identifying the progenitor
star from HST archival images, high-resolution ground based images \citep{vandyk, b43a}, and the
detection of a light echo \citep{b44b}. \cite{b16} estimate an expansion
velocity $\sim7000$\vunit\ from \pion{He}{i} 5876\AA, 6678\AA\ lines.

{There has been some uncertainty regarding the classification of SN~2016adj.
While consensus seems to lean towards Type IIb,
reports have variously suggested it as Type II \citep{b47}, Type Ib \citep{b39}, Type IIb
\citep{b16,b44}.} {It is classified as IIb by \cite{holoien} and we assume this classification here
(see also Section~\ref{IIb}).
Being a bright SN at outburst, we anticipate that it would have been extensively observed, and that a 
light curve should become available in due course which will help classify the object more robustly. 
This will also place our results and new knowledge in a better context.}

The supernova is located at RA 13$^{\rm h}$~25$^{\rm m}$~24\ddotsec11, Dec 
$-43^\circ$~$00'$~57\ddotarcsec50 (J2000) with a foreground
 2MASS \citep{2mass} star (2MASS 13252458--4300485) situated close to, and NW of, the supernova (see Fig.~\ref{cena}).
SN~2016adj is heavily reddened \citep[see][and references therein]{yang}. \citeauthor{yang} note the presence
of ``Diffuse Interstellar Bands'' -- commonly associated with interstellar dust -- in the optical
spectrum of SN~2016adj, and that emission from the SN
is strongly polarised; they also attribute the latter to significant reddening in the host galaxy.

\begin{table*}
\caption{Log of observations. \label{logobs}}

\centering
\begin{tabular}{ccccccccc}
\hline
UT       &  JD +   &  Days after  & IT($I\!J$)    & IT($J\!H$)   & IT($H\!K$)    &     Airmass        &  Airmass  &          Observatory\\
\scriptsize{YYYYMMDD} &  2457400             & discovery$^{a}$     &   (s)     &     (s)     &   (s)        &    (SN 2016adj)     & Std.$^{b}$&                \\

\hline
20160210 & 28.51 & 1.45&  760   &  760   &  760   &  2.744  &   2.707  &  Mount Abu\\
20160216 & 34.45 & 6.39  & 760   &  760   &  760   &  2.613  &   2.623  &  "\\
20160217 & 35.46 & 7.40  & 760   &  760   &  760   &  2.622  &   2.649  &  "\\
20160221 & 39.43 & 11.37 & 760   &  760   &  760   &  2.626  &   2.651  &  "\\
20160222 & 40.45 & 12.39 & 760   &  760   &  760   &  2.625  &   2.619  &  "\\
20160223 & 41.45 & 13.39 & 760   &  760   &  760   &  2.626  &   2.620  &  "\\
20160229 & 47.46 & 19.40 & 380   &  380   &  380   &  2.803  &   2.891  &  "\\
20160302 & 49.44 & 21.38 & 1330   &  1330   &  1520   &  2.684  &   2.638  &  "\\
20160311 & 58.37 & 30.31 &   ---   & ---     &  1140   &   2.631  &       2.616   &  "\\
20160312 & 59.42 & 31.36 & 1520   &  1330   &  1520   &  2.718  &   2.663  &  "\\
20160313 & 60.40 & 32.34 & 950   &  760   &  760   &  2.647  &   2.632  &  "\\
20160325 & 72.38 & 44.32 & 1140   &  2470   &  ---    &  2.712  &   2.624  &  "\\
20160327 & 74.37 & 46.31 & --- & ---  &  760   &  2.678   &       2.647  &       "\\
20160407 & 86.27 & 58.21 & 760   &  760   & ---     &  2.798  &   2.867  &  "\\
20160409 & 87.34 & 59.28 & --- &  ---    &  2090   &   2.720  &        2.793  &       "\\
20160410 & 88.30 & 60.24 & 760   &  760   &  1900   &  2.618  &   2.630  &  "\\
20160414 & 92.39 & 64.24 &      \multicolumn{3}{c}{\dotfill 1912$^{c}$ \dotfill}  &   2.273  &  2.247   &    IRTF\\
\hline
\multicolumn{8}{l}{(a) Date of discovery = 2016 February 08.56 UT  =  JD  2457427.063.}\\
\multicolumn{8}{l}{(b) The standard star for the IRTF observations was  HD 119430, for  Mt Abu SAO 224089.}\\
\multicolumn{8}{l}{(c) Integration time (IT) for the entire  spectrum.}\\
\end{tabular}
\end{table*}

\section{Observations}

\subsection{Mt Abu}
\label{abu}
NIR spectroscopy in the 0.85 to 2.4\mic\ region, at resolution $R\sim1000$,
was carried out with the 1.2\,m telescope of the
Mount Abu Infrared Observatory \citep{b1} using the Near-Infrared Camera/Spectrograph
equipped with a $1024\times1024$ HgCdTe Hawaii array.
SN~2016adj, at a declination of $\sim43^\circ$S, is a challenging target from
Mt. Abu (latitude 24.5${^{\circ}}$N) because the target never rises above $22.5^\circ$ during observations.

Spectra were recorded with the star dithered to two positions along the slit, with one or more spectra being recorded in both
of these positions. The coadded spectra in the respective dithered positions were subtracted from each other to remove sky and
dark contributions. The spectra from these sky-subtracted images were extracted and wavelength calibrated using a combination
of OH sky lines and telluric lines that register with the stellar spectra. To remove telluric lines from the target's spectra,
the latter was ratioed with the spectral standard SAO~224089
(Sp. type A3V, $T{_{\rm eff}} = 8720$~K),
from whose spectra the hydrogen Paschen and Brackett absorption lines had been removed.  The spectra were finally
multiplied by a blackbody at the effective temperature of the standard star. All data reduction was done using
IRAF\footnote{IRAF is distributed by the National Optical Astronomy Observatories,
which are operated by the Association of Universities for Research in Astronomy, Inc.,
under cooperative agreement with the National Science Foundation.}
tasks and self-developed IDL routines along the standard lines that we use for spectral reduction. Some earlier
spectroscopic results on SNe from Mt. Abu can be found in \cite{b23} and \cite{b29}, while several nova results
are presented in \cite{b1} and references therein.

\subsection{IRTF}
A spectrum of SN~2016adj was obtained using the 3\,m NASA Infrared Telescope Facility (IRTF) on 2016 April 14,
covering the 0.78 to 2.5\mic\ region. This spectrum was obtained using SpeX \citep{b36} in the cross-dispersed mode, using
the $0.5'' \times 15''$ slit ($R = 1200$) and a total integration time of 1912\,s. The SpeX data were reduced and
calibrated using the Spextool software \citep{b13}, and corrections for telluric absorption were performed using the
IDL tool xtellcor \citep{vacca}. 

\begin{figure*}
\includegraphics[width=16cm,keepaspectratio]{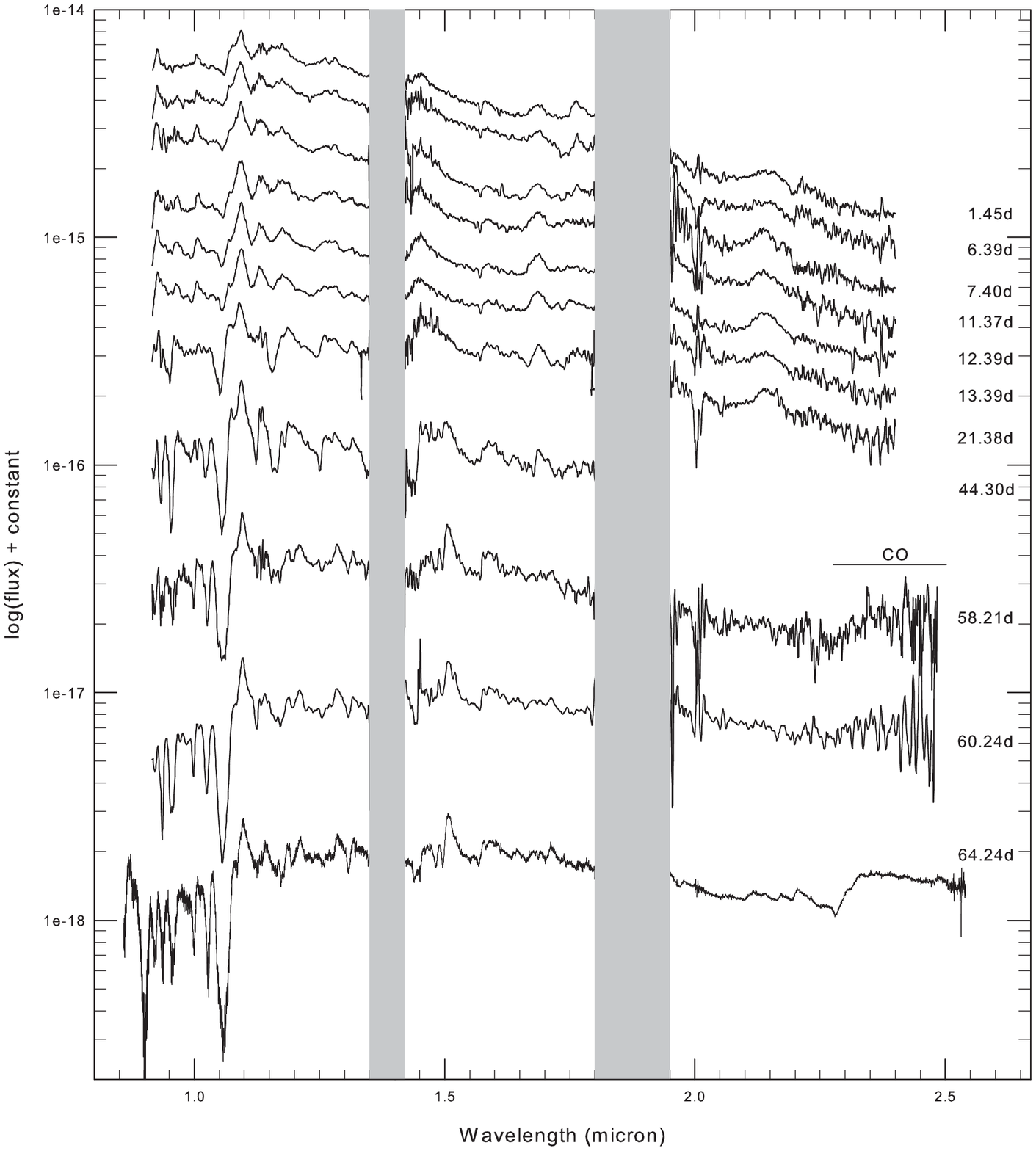}
\caption[]{NIR spectra of SN~2016adj are shown with the days after discovery indicated.
The bottom-most spectrum (+64.24d) is from the IRTF, the remainder are from Mt Abu. 
The Mt Abu spectra have been smoothed by a five point running average. 
Identification of the lines is given in Table~\ref{LineIDt} and Fig.~\ref{LineID}. 
The most significant finding is that of first overtone CO emission, which is seen from +58d.
Vertical bars denote regions of poor atmospheric transparency.
\label{fig2}}
\end{figure*}

The log of all the observations is given in Table~\ref{logobs}.
The NIR spectra, from day 1 to day 64, are presented in Fig.~\ref{fig2};
line identifications are given in Table~\ref{LineIDt}.

 \begin{table*}
\begin{center}
\caption{Line identifications.\label{LineIDt}}
\begin{tabular}{cccc} \hline\\
$\lambda$ (Obs)& ID    & $\lambda$ (Rest) & Transition\\
($\mu$m)       &       &    ($\mu$m)      & $\ell-u$ \\ \hline
0.9226         & Pa-9  & 0.92315 &   3--9 \\
0.9258         & \pion{Mg}{i} & 0.92583   &  $^1$D$-^1$F$^o$  \\
0.9658         & \pion{C}{i}  & 0.96057, 0.96234 & $^3$P$^o-^3$S  \\
1.0049         & Pa-7 & 1.00521 & 3--7 \\
1.0686         & \pion{C}{i}  & 1.06860, 1.06882, 1.06942 & $^3$P$^o-^3$D  \\
1.0831         & \pion{He}{i} & 1.08321,  1.08332, 1.08333 & $^3$S$-^3$P$^o$   \\
1.0938         & Pa-$\gamma$ & 1.09411 & 3--6\\
1.175          &  C blend & 1.17533 & $^3$P$^o_{3/2}$ -- $^3$P$^o_{5/2}$ \\
               &          & 1.17515,   1.17565, 1.17580  & $^3$D$-^3$F$^o$ \\
1.1287         & \pion{O}{i} &  1.12894, 1.12895, 1.12900, 1.12901 &  $^3$P$-^3$D$^o$\\      
               &             &  1.12902, 1.12904, 1.12982, 1.13008 &      \\
1.1330         & \pion{C}{i}  &  1.13334 &  $^1$P$-^1$D$^o$  \\
1.2818         & Pa-$\beta$ & 1.28216 & 3--5 \\
1.3164         & \pion{O}{i}  &  1.31675, 1.31685, 1.31687 &  $^3$P$-^3$S$^o$ \\
1.4543         & \pion{C}{i} &  1.44240  &  $^3$P$-^3$D$^o$ \\
               &             &  1.45465  &  $^1$P$^o-^1$P \\
1.5040         & \pion{Mg}{i} &   1.5025, 1.5040, 1.5048 & $^3$S$-^3$P$^o$ \\
1.6890         & \pion{C}{i} & 1.68675   &  $^1$P$-^1$P$^o$ \\    
               &             & 1.68950   &      $^1$D$-^1$F$^o$    \\
2.1656         & Br-$\gamma$ &2.16612  &  4--7  \\
2.2056         & \pion{Na}{i}  & 2.2056, 2.2084 & $^2$S$-^2$P$^o$  \\
2.266          & \pion{Ca}{i} & 2.261, 2.263, 2.266 & $^3$D$-^3$F$^o$ \\
\hline
 \end{tabular}
\end{center}
\end{table*}

\begin{figure}
\includegraphics[width=8cm,keepaspectratio]{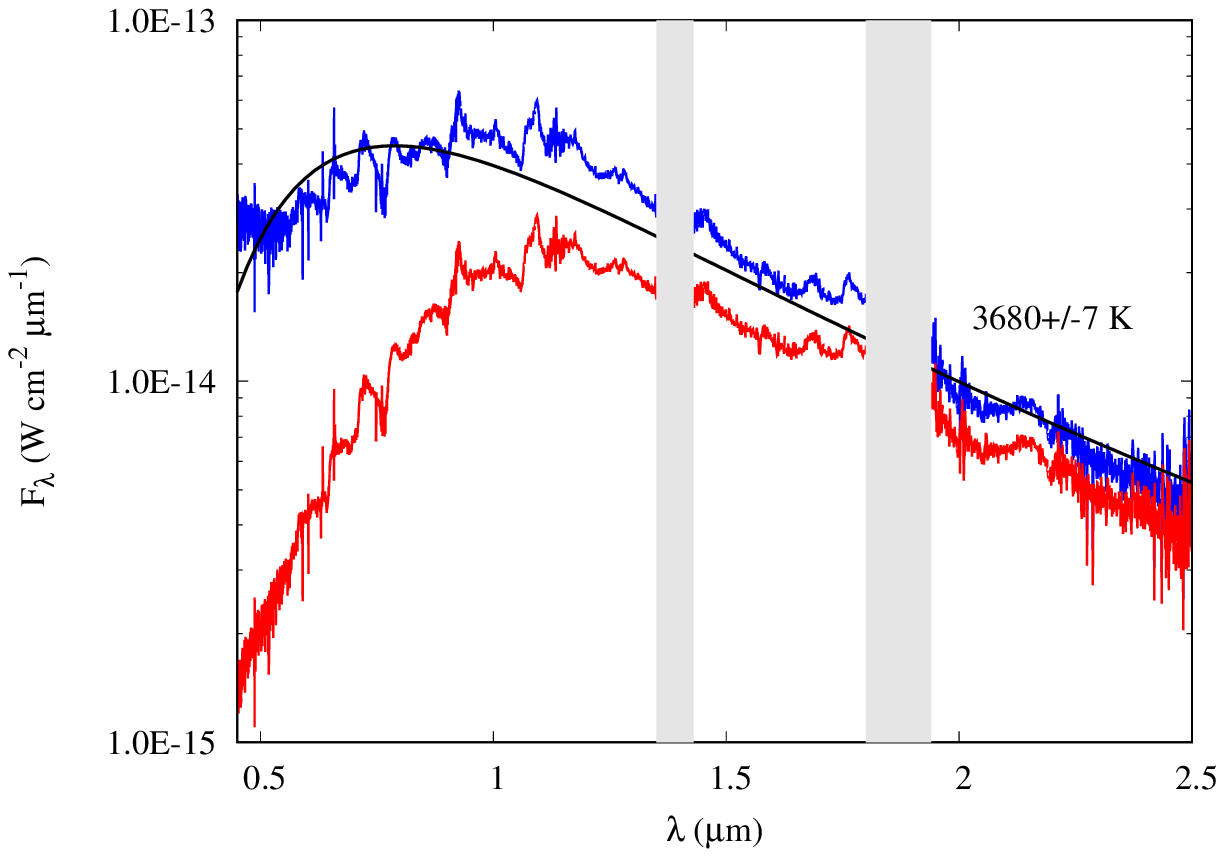}
\includegraphics[width=8cm,keepaspectratio]{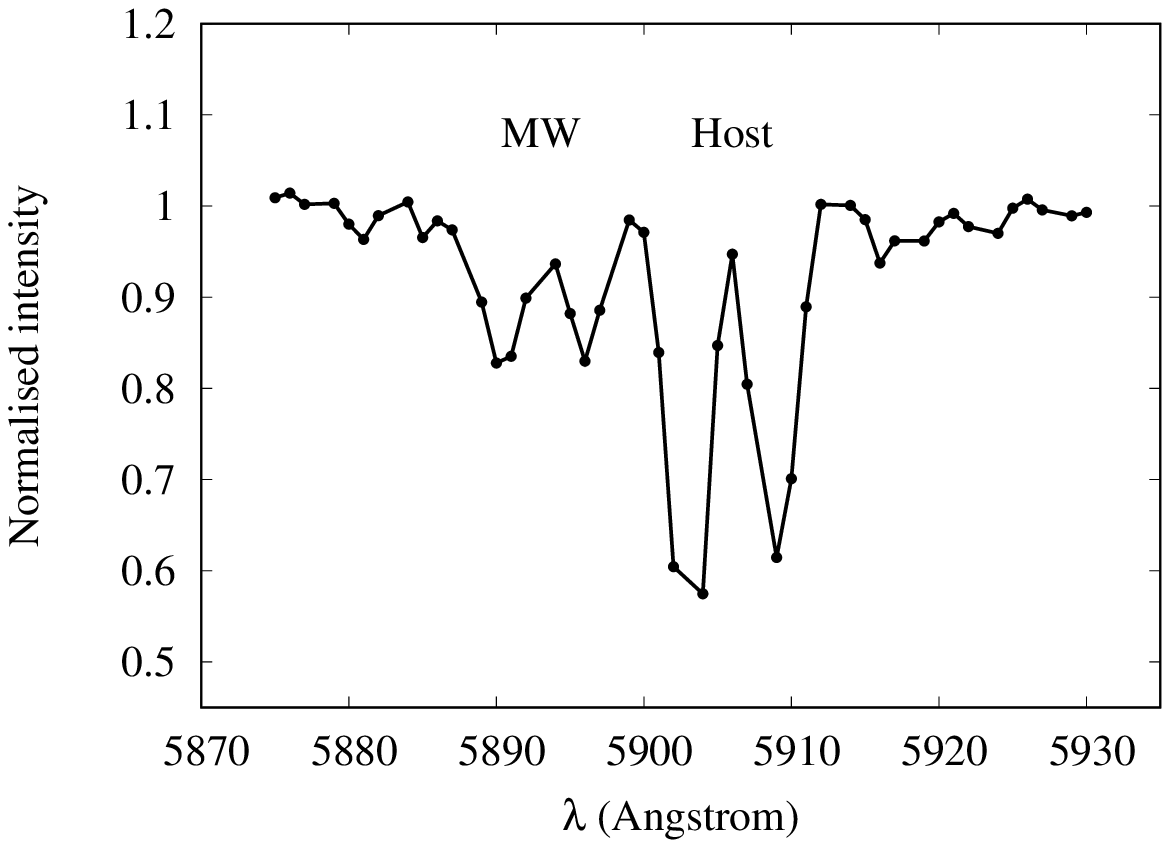}
\caption{Upper panel: Red curve, observed  spectrum for 2016 February 9, after combining the optical
(Thomas et al. 2016) data and the Mt. Abu NIR data.
Blue curve, spectrum after dereddening as discussed in text; the black curve is a 3680~K blackbody.
Vertical bars denote regions of poor atmospheric transparency.
Bottom panel: the \pion{Na}{i} D1, D2 lines in the SN spectrum using the Wiserep spectrum from
Thomas et al. (2016). The  Na\,D components from the host galaxy and the Milky Way are marked.
See text for details.}
\label{fig1}
\end{figure}

\section{Reddening and SED} \label{reddening}

An optical spectrum of SN~2016adj, covering the 3500--9800\AA\ region at 1\AA\ resolution,
was obtained by \cite{b44} on 2016 Feb 09.762 with the Wide Field Spectrograph
(WiFeS); the spectrum is available
on WISERep\footnote{https://wiserep.weizmann.ac.il/}. 

We present a combined optical \citep{b44} and NIR spectrum of 2016 February~9
-- before CO formation -- in Fig.~\ref{fig1}.
Merging of the WiFes and Mt~Abu spectra is a straightforward process, achieved  by matching the
continuum levels and emission features in the  overlapping region that both spectra
share around the 0.9\mic\ micron region.

In the optical spectrum \citep{b44} the \pion{Na}{i} D1 and D2 components are resolved, 
both from the Milky Way and the host galaxy, as shown in Fig.~\ref{fig1}.
The measured equivalent
widths (EW) of the D1 and D2 lines from the host galaxy are 1.34\AA\ and 1.45\AA\ respectively, 
and from the Milky Way are 0.58\AA\ and 0.79\AA\ respectively,
with errors typically of $\pm10$\%. 
We use the EWs to determine the reddening to SN~2016adj,
using the reddening versus EW relationships of \cite{b27}, 
based on the \cite{b33} data.

For the Milky Way, we get an average $E(B-V){_{\rm MW}} = 0.30\pm0.03$;
the same value is obtained from the reddening maps of \cite{b31}.
The EWs for SN~2016adj are slightly larger than the
calibrated range of values in the relationships of \citeauthor{b27}.
With this caveat, and assuming that the (Galactic) 
relationship between $E(B-V)$ and \pion{Na}{i} equivalent widths applies
to the dust lane of NGC~5128, we follow \citeauthor{b27} to
find that $E(B-V)$ values from the D2 and D1 lines are $0.55\pm0.05$ and $0.64\pm0.06$ respectively,
leading to an average $E(B-V){_{\rm host}}  = 0.60\pm0.06$.

For the Milky Way, we assume a ratio of total-to-selective extinction $R_V=3.1$. For
the NGC~5128 dust lane the value of $R_V$ is significantly lower, $2.57^{+0.23}_{-0.21}$, determined
by \cite{patat} on the basis of the wavelength-dependence of interstellar polarisation of SN~1986G
in the host galaxy; we assume the \citeauthor{patat} value here. We use the dereddening formalism of \cite*{CCM}.
Dereddening the SN~2016adj spectrum first for Galactic extinction, and separately
for extinction in the host galaxy, gives the spectral energy distribution (SED) in Fig.~\ref{fig1}.

After dereddening, the location of the continnum was estimated by eye, and
a blackbody was fitted to the combined optical-NIR continuum in Fig.~\ref{fig1}.
The fit yielded a T${_{\rm bb}} = 3680$~K, with a formal error of $\pm7$~K; 
however this error is a gross underestimate,
and is very dependent on the correct identification of
regions free of spectral (including P-Cygni) features.
Nevertheless, it is evident from Fig.~\ref{fig1} that the dereddened SED peaks around $0.7-1$\mic, 
suggesting that the photospheric temperature must be of this order.

Clearly, the photospheric temperature at early times for SN~2016adj is very much on the low 
side compared to the range of 6000--10000~K displayed by several stripped envelope SNe early after discovery.
For example, Figure 4 of \cite{b45} shows several stripped envelope SNe 
at or around discovery; these have $(B-V)\simeq0$, equivalent to spectral type $\sim$A5--B5, 
or effective temperature in the range 
$\sim8500-15000$~K. The low T${_{\rm bb}}$ in the case of SN~2016adj possibly presages 
conditions that are conducive for the formation of CO, subsequently detected in this SN.

We note that the dereddened SED in Fig.~\ref{fig1} broadly resembles that computed
by \cite{dessart} for a Type IIb SN at 4~d after eruption (the earliest in their
compilation; see their Figure~10, model 3p65Ax1). This shows a cool ``photosphere'' (peaking
around 0.9\mic\ in $\lambda^2f_{\lambda}$) with H lines sporting prominent P-Cygni
profiles. Their model 6p5Ax1, describing a Type Ib SN \citep[Figure 11 of][]{dessart}, 
is lacking in H lines, seemingly ruling out this SN Type for SN~2016adj.


\section{First-overtone CO emission}
\label{CO}

From 2016 April 7 (day~58.21), the Mt Abu $K$-band spectra (Fig.~\ref{fig2}) began to show a
rise towards the red starting from around 2.29\mic, suggesting first overtone
CO emission. A subsequent spectrum from the IRTF, obtained on day 64.24, 
clearly confirms this (see Fig.~\ref{fig2}),
and shows emission from the $\Delta{\upsilon} =2$ vibrational rotational transitions, extending up to 2.5\mic\
(the 8--6 and 9--7 bandheads are expected at approximately 2.48\mic\ and 2.51\mic\ respectively).

As discussed in Section~\ref{adj}, the optical maximum of SN~2016adj was reached
between 3 and 8.56 February. Our first CO detection was made on 2016 April 7, i.e. 58--64~d
after maximum, 58.2~d after discovery, thereby making it the second earliest CO detection in a SN
after SN 2013ge (48~days).

The signal-to-noise ratio (125) in the CO emission region of the IRTF spectrum is,
with the exception of SN~1987A \citep{b40}, higher compared to many
of the earlier CO detections, including that in SN~2013ge \citep[see][]{drout}.
The quality of our data allows much more reliable modelling of
the CO parameters, and thus a better understanding of the early time CO chemistry.
However before discussing our modelling we first consider whether it is appropriate
to include $^{13}$CO in our modelling.

\subsection{The case for $^{13}$CO}

It is believed that most of $^{13}$C (along with $^{15}$N and $^{17}$O)
in the Galaxy is synthesised in nova eruptions occurring on both Carbon-Oxygen and ONe core white dwarfs
\citep{starrfield,jose98}. These studies predict  significant $^{13}$C enrichment, which is confirmed
observationally \citep[see][who compile the $^{12}$C/$^{13}$C ratio estimated in the 10 novae which have shown
first overtone CO emission]{banerjee}. 

However the question has been raised whether CC SNe
might also contribute to Galactic $^{13}$C \citep{nittler}. These authors showed that  Si, C, N, Mg-Al, Ca
and Ti isotopic  data for three micron-sized presolar SiC grains from the Murchison meteorite had  very low $^{12}$C/$^{13}$C
ratios ($4.02\pm0.07$, $6.48\pm0.08$ and $5.59\pm0.13$ respectively), similar to that expected from nova grains.
However other isotopic signatures in  these grains ($^{28}$Si, $^{49}$Ti and $^{44}$Ca excesses,
and a very high inferred $^{26}$Al/$^{27}$Al ratio) indicated that they had  in fact formed in a SN
and not a nova.  

A similar conclusion is reached by \cite{nliu} from a  multi-element isotope analysis of their C2
class of  pre-solar SiC grains, which show large enrichments in both $^{13}$C and $^{15}$N.
Four such grains in their sample displayed low values ($1.9\pm0.03$, $6.4\pm0.08$, $1.6\pm0.02$
and $1.0\pm0.01$) for the  $^{12}$C/$^{13}$C ratio, apparently suggesting an origin in novae.
However, other isotopic signatures (excesses in $^{29}$Si ,$^{30}$Si, $^{50}$Ti)
clearly favoured an origin from neutron-burst nucleo\-synthesis in CC SNe.
\citeauthor{nliu} thus conclude that the enrichments of $^{13}$C and $^{15}$N seen in their C2 grains
point towards a CC SNe origin, and strongly suggest the occurrence of explosive H burning in the
He shell during CC SN explosions. 

{The H is introduced into the He shell by mixing, as in (for example) the  model developed  by 
\cite{pignatari} and discussed extensively in \cite{nliu}.
In view of the fact that H lines were strongly seen in early spectra of SN~2016adj 
(see Table~\ref{LineIDt} and Fig.~\ref{fig2}) --
indicating that it had a H envelope at outburst -- the ingestion of H into the He shell 
via mixing is quite plausible. \citeauthor{pignatari} have studied the nucleo\-synthesis 
impact of the ingestion of modest amounts of H into the He shell in the pre-SN stage,
i.e. before the SN shock reaches these layers. 
In some of their models (see model 25T-H, for example; top panel of their Figure 1) 
significant enhancement
of $^{13}$C production is found in an oxygen rich region \citep[designated by][``the O/nova zone'']{pignatari}. 
We speculate that the $^{13}$C$^{16}$O could likely form here, and avoid subsequent destruction by He$^+$ in this region 
\citep{lepp} if the  He$^+$ gets rapidly destroyed by charge exchange reactions with other metals (e.g C, O, Si)  
as was proposed in the case of SN~1987A \citep{lepp}. 
Such metals  are present in the O/nova zone as a consequence of mixing between layers \citep{pignatari}.}

These studies raise the possibility that grains previously attributed to novae might 
in fact have originated in Type II SNe; indeed
the presence of SiC dust -- which provided the rationale for including $^{13}$CO in
our analysis
--  in a young supernova remnant is suggested by \cite{rho18a}.
$^{13}$C production in a CC SN seems possible, and $^{13}$CO is therefore
included as a component in our modeling of the CO emission in SN~2016adj.
As our IRTF data were obtained at high signal-to-noise, they offer a unique opportunity to test 
the important hypothesis that 
CC SNe contribute significantly to the Galactic $^{13}$C budget. 

\subsection{Modelling the CO emission}

We consider two classes of models. The first class has  $^{12}$C$^{16}$O  only, whereas the second class
has both $^{12}$C$^{16}$O and $^{13}$C$^{16}$O in varying proportions. 
Only $^{12}$C$^{16}$O and $^{13}$C$^{16}$O are considered as likely contributing species to the CO emission;
the contribution of other isotopologues, such as $^{12}$C$^{17}$O, $^{14}$C$^{16}$O, is assumed to be negligible.

The CO emission has been analysed using the model  developed for nova V2615 Oph \citep*{das09},
in which strong first overtone CO emission was observed. This model was subsequently applied to three
other novae that showed CO emission, namely V496~Sct \citep{raj12}, V5584~Sgr 
\citep{raj15} and V5668~Sgr \citep{banerjee}.

The  CO is assumed  to be in thermal equilibrium characterised by the same rotational and vibrational
temperature \citep[as in the LTE model of][]{b40}; the populations of the different levels are
determined from the Boltzmann distribution.
The necessary ro-vibrational constants of CO  were taken from the National Institute of Standards and Technology data base
(http://physics.nist.gov/PhysRefData/MolSpec) and from \cite{benedict}.
Einstein-$A$ values were taken from \cite{goorvitch},  which lists these values for  all the ro-vibration
transitions of the first overtone band, up to $\upsilon = 20$ and $J = 149$ for seven CO isotopologues, including $^{13}$C$^{16}$O.
Other details of our model can be found in \cite{das09}. 
 
We also assume that the CO emission,
for both $^{12}$CO and $^{13}$CO, is optically thin. Should this not be the case then we expect
that our deduced CO mass will be an underestimate \citep[see for example][for the case of CO in SN~1987A]{LD}.
A full non-LTE treatment of the CO emission is beyond the scope of this paper.

\begin{figure} 
\includegraphics[width=7cm,keepaspectratio]{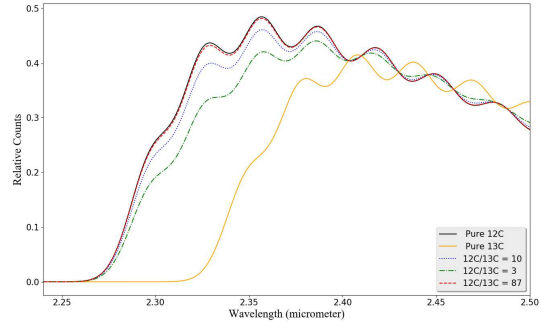}
\caption{{Models of first overtone CO emission for temperature 4600~K and velocity 3400\vunit.
Black curve: pure $^{12}$CO; 
{orange curve: pure $^{13}$CO; red, blue and green curves,} with $^{12}$C/$^{13}$C ratios indicated.
Note that $^{12}$C/$^{13}\mbox{C}=87$ is the solar value.}}
\label{12-13C}
\end{figure}

{We illustrate in Fig.~\ref{12-13C} the difference between the first overtone emission
from optically thin $^{12}$CO and $^{13}$CO, calculated using the models
by \citeauthor{das09}; we show calculations for 
temperature 4600~K and ejecta velocity 3400\vunit, typical of
the values we explore below. Fig.~\ref{12-13C} suggests that,}
{unless $^{12}$CO is dominant,
we can expect to be able to at least place constraints on the amount of $^{13}$CO.}

As discussed above, we have restricted our CO modelling to the 
high signal-to-noise IRTF spectrum of 14 April 2016 (64.24~d):
we exclude the  earlier Mt~Abu spectra.  The underlying  continuum
 was estimated  using a second order polynomial
 \[ f_\lambda = f_ 0 + a\lambda + b\lambda^2 \]
 for $f_\lambda$ in erg s$^{-1}$ cm$^{-2}$\mic$^{-1}$ and $\lambda$ in \mic; we find
 $f_0 = 1.58548395\times10^{-11}$~erg s$^{-1}$ cm$^{-2}$\mic$^{-1}$, 
 $a = -1.04831149\times10^{-10}$~erg s$^{-1}$ cm$^{-2}$\mic$^{-2}$ and 
 $b = 1.9003197\times10^{-11}$~erg s$^{-1}$ cm$^{-2}$\mic$^{-3}$;
 as shown in Fig.~\ref{COfits}, this closely resembles a 2550~K blackbody.
The form chosen for the continuum does not change the ratios of the band-heads, the quantity
that essentially determines the temperature.
 
Modelling over an extended  parameter space of 
temperature $T$,  velocity $V$  and $\alpha=^{12}$C/$^{13}$C
establishes that both the first and second class of model agree best
with the observed data with $T$ in the range  4000--5200~K,
and $V$ in the range 3400--3800\vunit. 

\subsubsection{Models with $^{12}$CO only}

\begin{figure*}
\includegraphics[bb=106 104 512 429,width=16cms,height=12.8cm,clip]{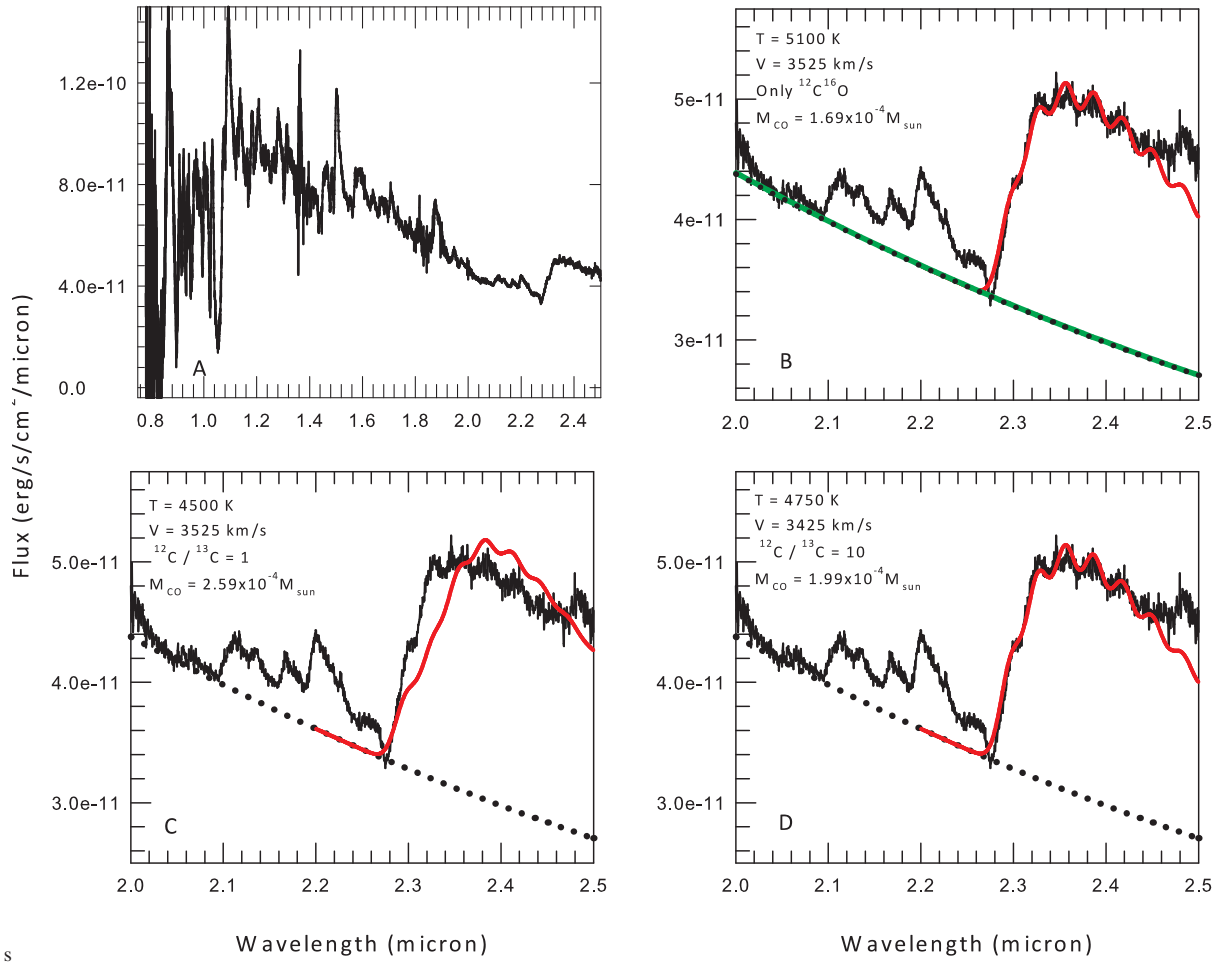}
\caption[]{Panel A shows the entire spectrum on day 64 after dereddening, as described in Section~\ref{reddening}.  
Panels B and D show fits (red curves) for a pure $^{12}$CO composition and a mixed ($^{12}$CO + $^{13}$CO) composition 
respectively, with the fit parameters indicated within. 
Panel C shows that, for $^{12}$C/$^{13}$C values less than 3, the quality of the fits are poor as the
contribution from $^{13}$CO begins to dominate. An example
of this is given for $^{12}\mbox{C}/^{13}\mbox{C} = 1$, where a good fit was not attainable for any 
combination of temperature and velocity. The black dotted lines show  the local continuum that has been 
adopted using a second order polynomial of the form 
$f_\lambda = f_ 0 + a\lambda + b\lambda^2$; this is found to closely match a blackbody
continuum at 2550~K, which is shown in green in panel B.}
\label{COfits}
\end{figure*}

Some  model plots, within this parameter subspace, are shown in
Fig.~\ref{COfits}, in which Panel A shows the entire spectrum.
Panel B shows the best fit for a gas composed entirely of $^{12}$CO. 
The goodness  of fit for the model spectrum to the observed spectrum was determined
by minimising the  reduced $\chi^2$ value.
In the observed spectrum the pure $^{12}$CO features  comprise the  knee at $\sim2.2998$\mic\
(which corresponds to  the $\upsilon = 2-0$  band-head of $^{12}$CO) and the $3-1$ band at 2.3227\mic.

We should note that underlying atomic lines  undermine the modelling. 
Some of these include  the suspected presence of a line at 2.489\mic,
which leaves a large residual in the fit at that point (see Fig.~\ref{COfits}).
This feature, likely due to  \pion{C}{i} 2.4894, 2.4901\mic,
also appears to be present in the spectra of SN~2013ge \citep{b17}.
There is also a puzzling absorption-like  feature at the foot of the CO emission at 2.275\mic,
also seen in SN~2004dj at 137~d post-explosion \citep[see Figure~16 of][]{b17}.

The best formal  fit for the pure $^{12}$CO case is obtained for a temperature of 5100~K 
and a velocity of 3525\vunit.
This is significantly lower  than the velocity estimated from optical \pion{He}{i} lines
\citep{b16}, and from the profile of the IR H lines (see Section~\ref{IIb}).
This emphasises the fact that there must be a degree of separation of the low 
excitation (CO-bearing) and high excitation (He-bearing) regions of the ejecta (see Section~\ref{COdust}).

While the  temperature of 5100~K for a pure $^{12}$CO  composition may seem higher 
than generally reported in the literature, it is a plausible value. 
A temperature of 5500~K was estimated in the Type IIP supernova SN~2004dj by \cite{sharp2}. 
Moreover, CO first overtone bands are seen in the spectrum of K type stars
\citep[e.g][]{kleinman} even up to spectral class  K0, which have a $T_{\rm eff} = 5280$~K;
the bands therein are more pronounced in giants and super giants compared to dwarfs.

\begin{table*}
\begin{center}
\caption{Estimated parameters of the CO emission on day 64.\label{CO-tab}}
\begin{tabular}{ccccc}
\hline\\
Composition     & $^{12}$C/$^{13}$C & T(K) & Velocity   & $M_{\rm CO}$     \\
     &           &     & (\vunit) & ($10^{-4}$\Msun)  \\
\hline                       \\
$^{12}$CO only       & --- & $5100\pm200$  & $3500\pm50$ &   $1.69\pm0.1$    \\
&&&&\\
$^{12}$CO + $^{13}$CO & $>3$& $4600\pm400$ & $3400\pm150$& $2.1\pm0.4$ \\ \hline
\end{tabular}
\end{center}
\end{table*}

\subsubsection{Models with $^{12}$CO and $^{13}$CO}

Models with $^{13}$CO  included in the composition are found to
fit the data satisfactorily, provided $\alpha>3$; 
our data place no tighter constraints than this
(as above we assess goodness  of fit by minimising the  reduced $\chi^2$). 
One of the satisfactory fits at
$\alpha = 10$ is shown in panel D of Fig.~\ref{COfits}. For $\alpha<3$,
the quality of the fits deteriorates rapidly, as the contribution from $^{13}$CO begins to dominate. 
An example of this is given in panel C for $\alpha = 1$, where a good fit was not attainable for 
any reasonable combination of temperature and velocity. This implies that the presence
of $^{13}$CO in the observed emission cannot be ruled out, but if it is present 
the $^{12}$C/$^{13}$C ratio must be $\gtsimeq3$
\citep[cf. the solar value of $\sim87$;][]{asplund}. Interestingly this conclusion is similar to that
drawn by \cite{matsuura17} for SN~1987A, who place a limit $^{12}$CO/$^{13}$CO$ > 21$ from their 
ALMA observations of the latter object.
However, we note here that the present work seems to be the first attempt to look for evidence
of $^{13}$C in a SN  using the first overtone CO emission; this has been made possible by the
high signal-to-noise in our IRTF spectrum.

As the criterion for C2 grains in particular is that $^{12}\mbox{C}/^{13}\mbox{C}<10$ \citep[see][]{nliu},
this shows that  CC SNe  as  sources of the $^{13}$C enrichment seen in certain  meteoritic
grains can not be ruled out. On the other hand, it seems unlikely that some of the extremely 
$^{13}$C-enriched C2 grains in the \citeauthor{nliu}  sample, with
$^{12}$C/$^{13}$C close to unity, could have been synthesised in a SN~2016adj-like SN. 

Our overall conclusion for models that include $^{13}$CO is that the observed CO emission 
is satisfactorily modelled  and characterised
with the following range of parameters:  $\alpha>3$,  temperature 
$T = 4600\pm400$~K, and
velocity $= 3400\pm150$\vunit, which lead to an estimated  CO mass $M_{\rm CO}=
2.1[\pm0.4]\times10^{-4}$\Msun.
These results are summarised in Table~\ref{CO-tab}.

\subsubsection{LTE and non-LTE}
\label{mind}

We should consider whether non-LTE calculations could yield results that are 
significantly different from those obtained assuming LTE. 
\cite{wliu} presented both LTE and non-LTE analyses for the spectra of SN~1987A 
obtained by \cite{b40}.  In the earliest 
spectrum, taken 192~d after discovery, the non-LTE fit is,
as these authors acknowledge, no better than the LTE fit; indeed 
the non-LTE fit is clearly  less satisfactory. It is only in the later spectra, 
between days 255 to 377, when the density falls and the  gas becomes optically thin that the non-LTE
models fare  significantly better than the LTE. With this is mind, and drawing a parallel between LTE and 
non-LTE values of the $T$ and $V$ values presented in Tables 1 and 2 of \cite{wliu}, 
we conclude that our estimated temperature and velocity values are not that different from
what a non- LTE calculation would have  given.
{However, given the differences in the deduced 
CO masses for SN~1987A  for the cases of LTE and non-LTE,
using the same datasets \citep{wliu}, it is likely that 
our LTE mass estimate may be  underestimated by a factor of  $\sim4$
\citep[see also Figure~3 of][]{sarangi2}}.

\subsection{Comparison of the CO modeling with previous studies}

\begin{table*}
\begin{center}
\caption{{Constraints on emergence of CO emission in SNe. 
Data for pre-2013 SNe from Gerardy et al. (2002), later SNe from Sarangi et al. (2018).\label{CO-SN}}}
\begin{tabular}{ccccc}  \hline\\
SN       & Type    &  Latest           & Earliest       & Reference\\ 
         &         & non-detection$^a$ (d) & detection$^a$ (d)  &             \\   \hline
1987A    & IIp Pec &    18             &    110?$^b$(192)         &  \cite{meikle0}          \\
1995ad   &  IIP    &    ---            &    105         &  \cite{SL}          \\
1998dl   &  IIP    &    ---            &    152         &  \cite{b42}          \\
1999em   &  IIP    &    ---            &    170         &   \cite{b42}         \\
2002hh   &  IIP    &    137$^c$        &    200         & \cite{pozzo} \\
2004dj   &  IIP    &    ---            &    106         &  \cite{kotak5}        \\
2004et   &  IIP    &      64           &    300         &  \cite{kotak9}   \\
2005af   &  IIP    &    ---            &     194          &  \cite{kotak6}  \\
2017eaw  &  IIP    &    53             &    107         & \cite{rho18}  \\ 
         &         &                   &                & \\
1998S    &  IIn    &    44             &    109         &  \cite{b12}         \\         
         &         &                   &                & \\         
2011dh   &  IIb    &    ---            &    89          &   \cite{b11}         \\         
2016adj  &  IIb    &    21          &    58       & This work  \\          
&         &                   &                & \\         
2000ew   &  Ic     &    39             &      97        &   \cite{b15}         \\
2007gr   &  Ic     &    46.9           &    70.8        &  \cite{b17}          \\
2013eg   &  Ib/c   &   ---             &    48          &  \cite{drout}          \\
\hline
\multicolumn{5}{l}{$^a$Days from discovery.}\\
\multicolumn{5}{l}{$^b$We have reassessed the CO detection in SN 1987A for day 110,
 reported by \cite{meikle0}; the data for this time seem to show}\\
\multicolumn{5}{l}{~~~the CO band-head; there is a clear detection at day 192.}\\
\multicolumn{5}{l}{$^c$The spectra of SN 2002hh is noisy so that it could contain CO band-head.}\\
\end{tabular}
\end{center}
\end{table*}

It is instructive to compare our estimates of the CO temperature and velocity with corresponding
estimates of the CO parameters in other SNe. 
A summary of CO detections in SNe before $\sim2002$ is given in Table~1 of \cite{b15};
{a summary of later detections is given in Table~1 of \cite{sarangi}.
These are gathered together here in Table~\ref{CO-SN}, together with more recent detections;
the SNe in this Table are grouped acccording to Type.}

{We first discuss how the CO parameters in SN~2016adj compare with those
of other SNe in which CO has been detected, and follow this with the implications for 
our understanding of CO formation in SNe.}

\subsubsection{SN 1987A}

The CO emision in SN~1987A was the best observed and most 
thoroughly analysed. In this object, substantial variation between 192~d to 377~d is seen 
in these parameters with time, especially in the temperature, along with variations also 
arising depending on the choice of the model used (LTE versus non-LTE). 
The reader is referred to the temperature, mass and velocity values in Tables 1 and 2 of \cite{wliu}, 
which analyses the data of \cite{meikle0}. \cite*{GWS}, analysing the SN~1987A data, 
suggest that on day 192, a temperature closer to 4000~K,  as also estimated by \cite{sharp}.

\subsubsection{SN 2004dj}

Detailed modelling of the fundamental CO emission spectra of the Type IIP supernova SN~2004dj 
has also been reported by \cite{kotak5}, who also detect the first overtone emission but do not model it. 
Comparison between observed and model profiles for the fundamental band of CO on day 130 
shows that the best agreement is obtained
with $T = 5500$~K; this is even higher than our estimate in this study. 
While these authors do not mention the velocity of their 
best fit model, it is implicitly presumed to  lie in the range of $2000-4000$\vunit.

\subsubsection{SNe 2007gr, 2000ew}

While \cite{b17} and \cite{b15} did not carry out detailed modelling of the 
CO emission seen in SNe~2007gr  and 2000ew  respectively, they
estimate the temperature and velocity
by  overplotting their CO profiles on the  SN~1987A profiles, 
and comparing the shapes and slopes of various band-heads. \citeauthor{b17}  conclude that, 
on day 137.7, the temperature of the CO emitting region in 
SN~2007gr was greater than 2000~K, and the expansion velocity was in 
the range 1800--2000\vunit. \citeauthor{b15}
conclude that,  on day 97,  the temperature was at least 2000~K in 
the CO emitting region, and that the CO has a 
velocity similar to that of SN~1987A, $\sim2000$\vunit. 
However the spectrum of \citeauthor{b15} is of low signal-to-noise.

\subsection{Timescale of CO formation in SNe}

Our early detection of CO in SN~2016adj suggests that there is a range
of time-scales over which CO and dust form in the ejecta of Type II SNe. A number of parameters,
including hydrogen mixing, the C/O ratio 
and metal depletion, could contribute to those timescales \citep{cherchneff1}.
Models of CO formation during SN eruption indicate that the amount of
CO increases from $\ltsimeq10^{-4}$\Msun\ at 100 days since the explosion,
and thereafter increases to as much as 0.1\Msun\ after $\gtsimeq1500$~days,
irrespective of SN progenitor \citep[see e.g.][]{sarangi2}.
Figure~10 of \cite{sarangi}, summarising data to  2013, shows only four CO detections
before day~100.

The  summary of CO detections according to SN type (Table~\ref{CO-SN}) suggests that, in Type II~P
SNe, most CO emission is detected for the first time shortly after 100 days, 
consistent with CO formation model of \cite{sarangi2,sarangi3}. Type~IIn are SNe
that show interaction with the circum-stellar medium (CSM), and SN~1998S is the only Type~IIn which 
shows behaviour similar to those of Type IIP. 
\cite{sarangi} show that dust formation associated with a CSM occurs only after
day~380, which may indicate that the CO and dust evolution
of Type~IIn SNe may not be very different from that of Type IIn.
However, a larger sample is required to draw any firm conclusion.

On the basis of just two examples (SN~2000ew and SN~2007gr), the 
CO formation time-scale for Type~Ic SNe appears to be much shorter than
that for  Type IIP SNe. The presence of ionised helium hinders CO formation 
\citep[][their Figure~7]{LD,sarangi2}, while Type Ic SNe -- with stripped helium -- may show
enhance CO formation.

A search of the literature indicates there is little theoretical work which addresses the formation 
of CO at very early times ($\ltsimeq50$~d or less), as observed in SN~2013eg or  SN~2016adj;
for example the \cite{sarangi2}  models begin simulations at 100~d. 
Thus it is difficult to place the early formation
of CO in SN~2016adj into a theoretical framework  
to compare observations and model predictions. In fact SNe such as SN~2013eg, SN~2016adj or SN~2010gr --
all of which display early emergence of CO --  could provide impetus for fresh  modelling to understand 
early CO formation, the amount of CO expected to be produced, and the location of the CO production zones.

\section{General characteristics of the NIR spectra}
\subsection{Line identification}
\begin{figure}
\includegraphics[width=3.0in,keepaspectratio,clip]{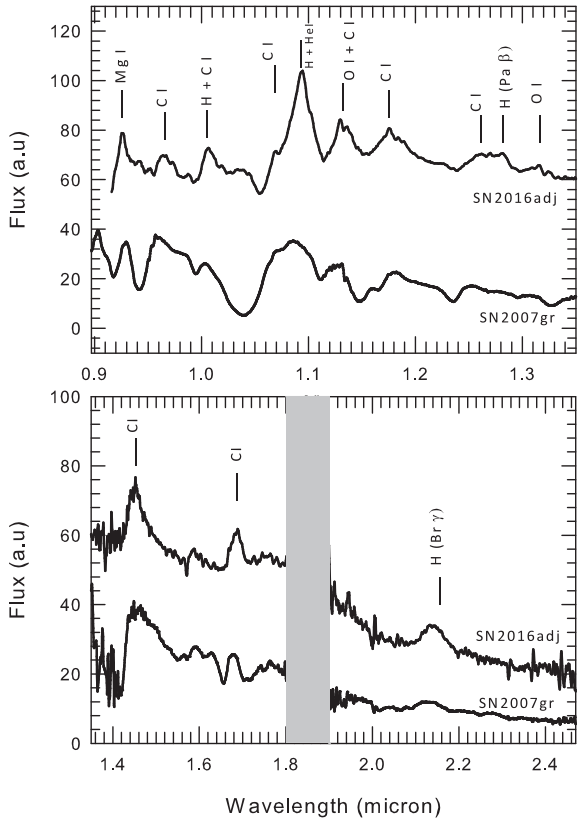}
\caption{Line identification of the spectrum of 2012 Feb 22 (+12d). 
The spectrum of SN~2007gr on +15.5d is also shown to show considerable similarity between both spectra.
SN 2007gr showed several carbon lines in its spectrum (see panel 2 of Figure 12, Hunter et al. 2009)
which are similarly seen here and marked.  \label{LineID}}
\end{figure}

The prominent lines seen in the spectra are identified in Fig.~\ref{LineID},
and listed in Table~\ref{LineIDt}.
It appears that we may  be seeing several carbon lines in our spectra.
We have compared our spectra with that of the carbon-rich Type Ic supernova  SN~2007gr on +15.5d 
\citep[][their Figure 12]{b17}. Both spectra show considerable similarity, and
most of the carbon features identified by \citeauthor{b17} are also seen in SN~2016adj.
We note that the C lines identified here are routinely encountered 
 in the spectra of novae, where they are much easier to identify because of their smaller line widths. 
It may be seen that the \pion{O}{i} feature at 1.1287\mic\ shows considerable fluctuation
from epoch to epoch. This is primarily because the line is situated in a window of 
very poor atmospheric transmission and small differences  in airmass between source and 
standard star can create large residuals and substantial noise in that region
 during the process of ratioing of the source and standard star spectra.
 Notwithstanding this, we note that the \pion{O}{i} 1.1287\mic\ is pumped by Ly-$\beta$ photons
{\citep[see e.g.][]{stritt,AGN2}},
 so the strength or absence of this feature might also be attributed
to a lack of velocity overlap between the lines.

\subsection{IIb or not IIb}
\label{IIb}

\begin{figure}
\includegraphics[width=3.0in,keepaspectratio,clip]{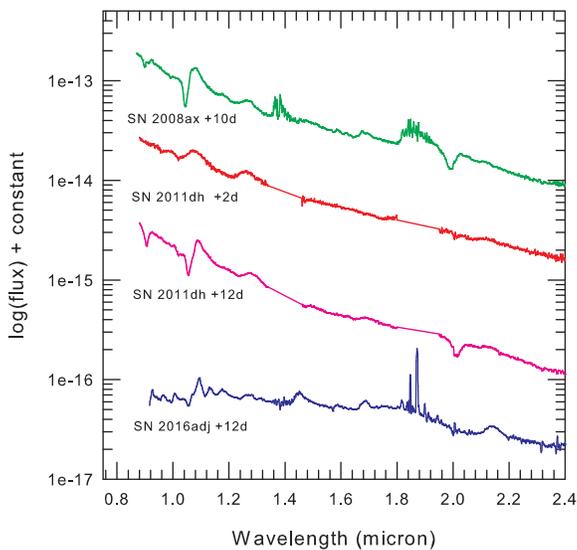}
\caption{The NIR spectra of two well-studied SNe of Type IIb, viz. SN~2011dh and SN~2008ax, are shown for comparison
with SN~2016adj. The strong and early development of the
 \pion{He}{i} 2.0581\mic\ line in the former SNe may be noted, while  the same line is absent in SN~2016adj. \label{sn-comp}}
\end{figure}

\begin{figure}
\includegraphics[width=8cm, clip]{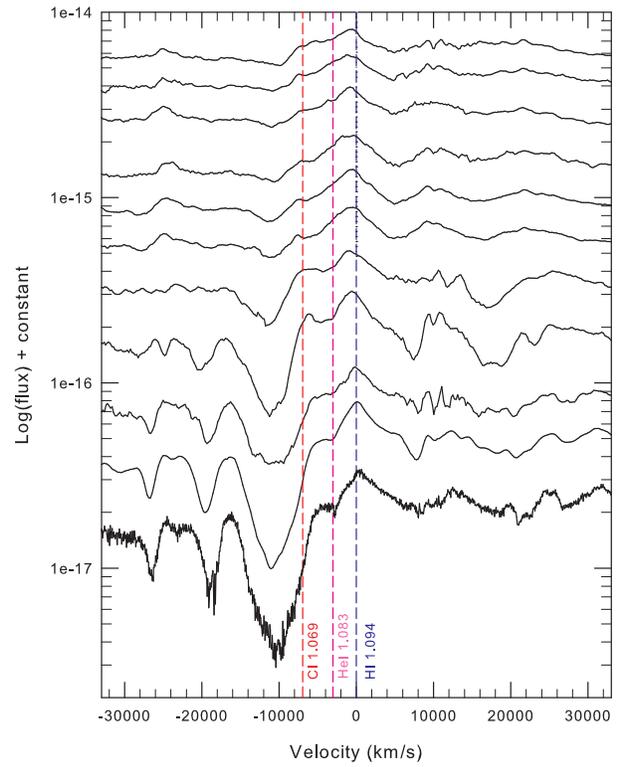}
\caption{The velocity evolution of the Pa-$\gamma$ 1.094\mic\ line is shown for all epochs.
 The expected positions of the \pion{C}{i} 1.069\mic\ and \pion{He}{i} 1.083\mic\ lines are marked.
 \label{fig4}}
 \end{figure}

Consistent with the traits of a SN IIb, SN~2016adj displays  prominent hydrogen lines from the outset.
Among these we unambiguously identify Pa-7 1.0049\mic, Pa-$\gamma$ 1.0938\mic, 
Pa-$\beta$ 1.2818\mic\ and Br-$\gamma$ 2.1656\mic.
As discussed in Section~\ref{reddening}, our data for day~1.45 broadly resemble the model SED of 
a Type IIb SN \citep[see][]{dessart}.

There is a  feature at 0.9226\mic\ that may be Pa-9, but it could be 
blended with \pion{Mg}{i} at a similar wavelength
($\lambda=0.92583$\mic; see Table~\ref{LineIDt}).
Although H lines are dominant,  a Type IIb supernova is expected to make  a transition from an 
initial hydrogen-dominated spectrum  to  the helium-dominated spectrum of a
SN Ib at later times. This can be seen in Fig.~\ref{sn-comp}, where  the spectra of SNe~2011dh and 2008ax, 
two well studied Type IIb SNe, are displayed, showing the strong and early development 
of the \pion{He}{i} 1.0831\mic\ and 2.0581\mic\ lines
\citep[see][for a large collection of spectra spanning early to late evolution]{b3, b9, b21, b45}.

In contrast there is little or no evidence for the presence of \pion{He}{i} 2.0581\mic\
in SN~2016adj, even up to 64~d after discovery.
Its absence indicates that SN~2016adj may be different from a typical SN IIb. 
After the 1.0831\mic\ line,
the 2.0581\mic\ line is the second strongest \pion{He}{i} line in the NIR region,
although the latter may be weak and is sometimes absent: this might be because
the upper level $^1$P$^o$ of the \pion{He}{i} 2.0581\mic\ line
has a strongly allowed transition to the ground $^1$S state, whereas there is no
such exit available for the upper $^3$P$^o$  level of the  1.0831\mic\ line.
[We note that (the \pion{He}{i} 1.8690\mic\ line is also strong, but lies in an
atmospheric window where the transmission is poor].

Towards the later stages (see, for example, the last two spectra of Fig.~\ref{fig2}),
there is a prominent shoulder forming on the blue wing of the 1.094\mic\ feature, which 
could result from the build-up of the \pion{He}{i} 1.0831\mic.
Attributing the peak of the 1.094\mic\ feature to H, the \pion{H}{i} expansion 
velocity measured between peak to absorption minimum,
is found to be 9150\vunit\ on day 1, which increases to $\sim11000$\vunit\ 
towards the end of the observations (see Fig.~\ref{fig4}).

The absence of the \pion{He}{i} 2.0581\mic\ line could indicate either that He is  deficient, 
or that  physical conditions are not suitable to excite He lines. 
If SN~2016adj is indeed a stripped envelope SN, the first layer of material 
that should be lost is hydrogen. Yet hydrogen is seen in all our spectra, implying that 
at least a portion  of it has been retained. It is therefore difficult to understand how  
the helium layer, which lies beneath the hydrogen layer, could have been so significantly stripped
that He should appear to be deficient. Rather, it appears more likely that the 
rather low photospheric temperatures of $\sim3680$~K, as indicated by the 
SED (Fig.~\ref{fig1}), are not conducive to collisional excitation or ionisation of He
(the $^{3}$P and $^{1}$P upper levels of the \pion{He}{i} 1.0830, 2.0581\mic\ lines have 
excitation energies 20.96 and 21.22~eV respectively).
Collisional ionisation followed by line formation through recombination,  appears even more problematic.
The role of the ambient radiation field in He line excitation  needs to be examined 
in detail but we do not attempt this here
\citep[see][for a discussion of the excitation of \pion{He}{i}]{maurer}.
However a cautionary note that emerges, specifically with respect to the taxonomic 
classification of SNe, is that the absence of He lines may not always mean that it is 
deficient or that the  He layer has been stripped.

{Also we note that \cite{sarangi2} studied the formation of molecules and dust clusters in the ejecta of 
solar metallicity, Type II supernovae using a chemical kinetic approach. The chemically
controlled simulation produces 
the evolution of molecules and small dust cluster masses from day 100 to day 1500 after explosion
for the stellar progenitors with initial mass of 12, 15, 19 and 25\Msun.  
Their simulaton demontrates that 
SNe with large progenitor masses tend to form dust efficiently, but are even more efficient at forming 
molecules; they find that the molecular component of the ejecta can be as large as $\sim$50\% of the total ejected mass. 
The decrease of He$^+$ content in the outermost ejecta zone by stripping (Liu et al. 2015)  enhance
CO formation. The Type IIb progenitor of SN~2016adj may be consistent with early CO formation,
compared with (for example) SN~1987A (Type II-pec).}

\subsection{Some $K$ band emission features of special interest}

Apart from the CO emission in the $K$ band discussed in Section~\ref{CO}, 
we identify the \pion{Na}{i} 2.2056, 2.2084\mic\ lines as
contributors to the prominent 2.2\mic\ blend seen in the CO modelling figures. 
These lines are known to effectively trace low temperature gas and invariably precede, 
in the case of novae, the formation of molecules and dust \citep{evans,b7}.
The reason for this can be understood from LTE calculations. At around 2500~K, 
only about 50\% Na remains neutral but by 3000~K, almost 99\% is ionised. 
These neutral \pion{Na}{i} lines therefore originate in relatively cool zones,
close to the temperature of $\sim1800$~K where the first dust condensates 
(carbon) are expected to form \citep[see e.g.][and references therein]{bianchi}.  
In the case of SN~2016adj observations taken in July-August 2016 confirm the 
formation of dust in this supernova through a significant NIR excess; this will discussed elsewhere.

Also notable is the likely absence  of the unidentified feature at 2.26\mic\ 
that was prominently seen in SN~1987A during CO emission,
as well as in a few other supernovae such as SN~1995ad \citep{SL}, SNe~1998dl and 1999em \citep{b42},
but absent in several others \citep{b15}. CO$^+$ was proposed as a potential carrier,
but was ruled out on the basis of detailed models of CO chemistry \citep{LD,GWS}.
Furthermore, as \cite{b15} point out, this feature did not evolve with the CO bands in SN~1987A,
remaining visible long after the CO emission has faded \citep{meikle}.
We propose that this feature could arise  from the combined presence of the \pion{Ca}{i} lines at
2.261\mic, 2.263\mic, 2.266\mic, 
which are routinely and prominently seen in the spectra of cool K and M stars
\citep[see e.g.][]{kleinman}. The low temperature environment in which these lines are excited
is compatible with similar conditions prevalent  in SN~2016adj,
as implied by the formation of molecules like CO.


\section{Conclusions}
Near-IR spectra, covering the first 64 days after discovery,  have been presented for the CC SN~2016adj 
in the nearby galaxy Centaurus A (NGC 5128). We detect first overtone CO emission on day 58, one of the earliest
CO detections in a SN. LTE modelling of the CO emission yields  temperature, velocity and mass  values 
of $5100\pm200$~K, $3500\pm50$\vunit\ and $1.69[\pm0.10]\times10^{-4}$\Msun\  respectively for a gas  
composed entirely of $^{12}$CO. 
If the CO has a component of $^{13}$CO also (in addition to  $^{12}$CO), then
our best fit models suggest the following range of values for the above parameters:
$4600\pm400$~K,  $3400\pm150$\vunit\ and $2.1[\pm0.4]\times10^{-4}$\Msun.
{However we should note that these masses may be lower limits,
as the CO may be optically thick (see Section~\ref{mind})}
The inclusion of $^{13}$CO was  motivated by the isotopic analysis  of pre-solar  SiC  meteoritic grains,
which show  large $^{13}$C enrichments and for which a CC SNe origin is strongly suggested \citep[e.g.][]{nittler}. 
Our analysis shows that the CO emission profile
is adequately explained by a pure $^{12}$CO composition but  does  not rule out the 
presence of $^{13}$CO provided it is such that  $^{12}\mbox{C}/^{13}\mbox{C}>3$.

We estimate the reddening to the object to be $E(B-V)= 0.90\pm0.09$, with the majority arising in the
host galaxy. A blackbody fit to the optical and near
IR data show that the photosphere has a low temperature of $\sim3680$~K at outburst.  
SN~2016adj has been proposed to be
a Type IIb supernova, which show an initial hydrogen-dominated spectrum but later, undergo
a transition to a helium-rich Type Ib spectrum. In SN~2016adj, hydrogen lines are prominent
initially, but the helium lines do not strengthen as expected. 
It is possible that this is a peculiar object,
or a SN making a rather slow transition from Type IIb to Ib.  
An unidentified  line at 2.26\mic\ line has often been seen in the spectrum
of SNe (e.g SN~1987A), which merges with the blue wing of the 2-0 band-head of CO. It is proposed that this
feature  arises from the \pion{Ca}{i} 2.260, 2.262, 2.265\mic\
transitions routinely seen in the spectra of cool stars.

\section*{Acknowledgments}

{We thank an anonymous referee for their comments.}
The research work at Physical Research Laboratory is supported by the Department of Space, Government of India.
RDG was supported by NASA and the United States Air Force.
MSC was supported under contract NNH14CK55B with the National Aeronautics and Space Administration.
TRG's research is supported by the Gemini Observatory, which is operated by the Association of Universities 
for Research in Astronomy, Inc., under a cooperative agreement with the NSF on behalf of the Gemini partnership: 
the National Science Foundation (United States), the National Research Council (Canada), CONICYT (Chile), 
Ministerio de Ciencia, Tecnología e Innovaci\'on Productiva (Argentina), and Minist\'erio da Ci\^encia, Tecnolog\'ia e 
Inova\c c{\~a}o (Brazil). 
JR acknowledges support from NASA ADAP grant NNX12AG97G for the study of SN dust.
We are very thankful to  A. Tokunaga for making the IRTF time available on short notice for this program.

\label{lastpage}

\begin{thebibliography}{99}


\bibitem[\protect\citeauthoryear{Asplund et al.}{2009}]{asplund} Asplund M., Grevesse N., Sauval A. J.,
       Scott P., 2009, ARAA, 47, 481

\bibitem[\protect\citeauthoryear{Banerjee \& Ashok}{2012}]{b1} Banerjee D. P. K., Ashok N. M.,  2012, BASI, 40, 243

\bibitem[\protect\citeauthoryear{Banerjee et al.}{2016}]{banerjee} Banerjee D. P. K., Srivastava M. K., 
          Ashok N. M., Venkataraman V., 2016,  MNRAS, 455, L109

\bibitem[\protect\citeauthoryear{Benedict et al.}{1962}]{benedict} Benedict W. S., Herman R., Moore G. E., Silverman S., 1962, ApJ, 135, 277

\bibitem[\protect\citeauthoryear{Bertoldi et al.}{2003}]{bertoldi03} {Bertoldi F., Carilli C. L., Cox P., 
Fan X., Strauss M. A., Beelen A., Omont A., Zylka R., 2003, A\&A, 409, L55}

\bibitem[\protect\citeauthoryear{Bianchi \& Schneider}{2007}]{bianchi} Bianchi S., Schneider R., 2007, MNRAS, 378, 973

\bibitem[\protect\citeauthoryear{Biscaro \& Cherchneff}{2014}]{biscaro} Biscaro C., Cherchneff I., 2014, A\&A, 564, A25

\bibitem[\protect\citeauthoryear{Cardelli, Clayton \& Mathis}{1989}]{CCM} Cardelli J. A., Clayton G.C., Mathis J. S.,
          1989, ApJ, 345, 245

\bibitem[\protect\citeauthoryear{Catchpole \& Glass}{1987}]{b2} Catchpole R.,  Glass I., 1987, IAUC 4457

\bibitem[\protect\citeauthoryear{Catchpole et al.}{1988}]{catchpole} Catchpole R., et al., 1988, MNRAS, 231, 75P

\bibitem[\protect\citeauthoryear{Cherchneff \& Dwek}{2010}]{cherchneff09} {Cherchneff I., Dwek E., 2009, ApJ, 703, 642}

\bibitem[\protect\citeauthoryear{Cherchneff \& Dwek}{2010}]{cherchneff10} {Cherchneff I., Dwek E., 2010, ApJ, 713, 1}

\bibitem[\protect\citeauthoryear{Cherchneff \& Lilly}{2008}]{cherchneff1} Cherchneff I., Lilly S., 2008, ApJ, 683, L123

\bibitem[\protect\citeauthoryear{Chornock et al.}{2011}]{b3} Chornock R., et al., 2011, ApJ, 739, 41

\bibitem[\protect\citeauthoryear{Clocchiatti et al.}{1996}]{b5} Clocchiatti A., Wheeler J. C., Benetti S., Frueh M., 1996, ApJ, 459, 547

\bibitem[\protect\citeauthoryear{Cushing, Vacca \& Rayner}{2004}]{b13} Cushing M. C., Vacca W. D., Rayner J. T., 2004, PASP, 116, 362

\bibitem[\protect\citeauthoryear{Das et al.}{2008}]{b7} Das R. K., Banerjee D. P.  K.,
Ashok N. M., Chesneau O., 2008, MNRAS, 391, 1874

\bibitem[\protect\citeauthoryear{Das, Banerjee \& Ashok}{Das et al.}{2009}]{das09} Das R. K., Banerjee D. P. K., 
        Ashok N. M., 2009, MNRAS, 398, 37

\bibitem[\protect\citeauthoryear{de Looze et al.}{2017}]{delooze} De Looze I., Barlow M. J., Swinyard B. M., 
             Rho, J., Gomez H. L., Matsuura M., Wesson R., 2017, MNRAS, 465, 3309

\bibitem[\protect\citeauthoryear{Dessart et al.}{2015}]{dessart} Dessart L., Hillier D. J., Woosley S.,
           Livne E., Waldman R., Yoon S.-C., Langer N., 2015, MNRAS, 453, 2189

\bibitem[\protect\citeauthoryear{Drout et al.}{2016}]{drout} Drout M. R., et al., 2016, ApJ, 821, 57

\bibitem[\protect\citeauthoryear{Egner et al.}{2010}]{egner} Egner S., Ikeda Y., Watanabe M.,
       Hayano Y., Golota T., Hattori M., Ito M., Minowa Y., Oya, S., Saito Y., Takami H., Iye M.,
       2010, SPIE, 7736, id. 77364V 

\bibitem[\protect\citeauthoryear{Ergon et al.}{2014}]{b9} Ergon M.,  et al., 2014, A\&A, 562, 17

\bibitem[\protect\citeauthoryear{Ergon et al.}{2015}]{b11} Ergon M., et al., 2015, A\&A, 580, 142

\bibitem[\protect\citeauthoryear{Evans et al.}{1996}]{evans} Evans A., Geballe T. R., Rawlings J. M. C., Scott A. D., 1996, MNRAS, 282, 1049

\bibitem[\protect\citeauthoryear{Fassia et al.}{2001}]{b12} Fassia A.,  et al., 2001, MNRAS, 325, 907

\bibitem[\protect\citeauthoryear{Fillipenko}{1997}]{b13a} Filippenko A. V., 1997, ARAA, 35, 309

\bibitem[\protect\citeauthoryear{Fox et al.}{2010}]{fox} {Fox O. D., Chevalier R. A., Dwek E., Skrutskie M. F.,
Sugerman B. E. K., Leisenring J., 2010, ApJ, 725, 1768}

\bibitem[\protect\citeauthoryear{Gall, Hjorth \& Andersen}{2011}]{gall} {Gall C., Hjorth J., Andersen A. C., 2011, A\&A Rev, 19, 43}

\bibitem[\protect\citeauthoryear{Gearhart, Wheeler \& Swartz}{Gearhart et al.}{1999}]{GWS} 
       Gearhart R. A., Wheeler J. C., Swartz D. A., 1999, ApJ, 510, 944

\bibitem[\protect\citeauthoryear{Gehrz}{1989}]{gehrz} Gehrz R. D., 1989, in IAU Symposium 135 Infrared Astronomy, p.~445,
   eds L. J. Allamandola, A. G. G. M. Tielens, Kluwer Academic Publishers, Dordrecht
       
\bibitem[\protect\citeauthoryear{Gehrz \& Ney}{1990}]{GN} Gehrz R. D., Ney A. P., 1990, Proc. Nat. Acad. Sci. (USA), 97, 4354
       
\bibitem[\protect\citeauthoryear{Gerardy et al.}{2000}]{b14} Gerardy C. L., Fesen R. A., Hoflich P.,  Wheeler, J. C., 2000, AJ, 119, 2968

\bibitem[\protect\citeauthoryear{Gerardy et al.}{2002}]{b15} Gerardy C. L.,  et al., 2002, PASJ, 54, 905

\bibitem[\protect\citeauthoryear{Gomez et al.}{2012}]{gomez} Gomez H. L., et al., 2012, ApJ, 760, 96

\bibitem[\protect\citeauthoryear{Goorvitch}{1994}]{goorvitch} Goorvitch D., 1994, ApJS, 95, 535

\bibitem[\protect\citeauthoryear{Holoien et al.}{2017}]{holoien} {Holoien T. W.-S., et al., 2017, MNRAS, 471, 4966}

\bibitem[\protect\citeauthoryear{Hounsell et al.}{2016}]{b16} Hounsell R. A.,  et al., 2016, ATel 8663

\bibitem[\protect\citeauthoryear{Hunter et al.}{2009}]{b17} Hunter D. J., et al., 2009, A\&A, 508, 371

\bibitem[\protect\citeauthoryear{Indebetouw et al.}{2014}]{indeb} Indebetouw R., et al., 2014, ApJ, 782, L2

\bibitem[\protect\citeauthoryear{Isaak et al.}{2002}]{isaak02} {Isaak K. G., Priddey R. S., McMahon
R. G., Omont A., Peroux C., Sharp R. G., Withington S., 2002, MNRAS,329, 149}

\bibitem[\protect\citeauthoryear{Jos\'e \& Hernanz}{1998}]{jose98} Jos\'e J., Hernanz M., 1998, ApJ, 494, 680

\bibitem[\protect\citeauthoryear{Kamenetzky et al.}{2013}]{kamenetzky} Kamenetzky J., et al., 2013, ApJ, 773, L34

\bibitem[\protect\citeauthoryear{Kiyota et al.}{2016}]{b19} Kiyota S., Shappe B. J., Stanek K. Z., Subo D., 2016, ATel 8654

\bibitem[\protect\citeauthoryear{Kleinman \& Hall}{1986}]{kleinman} Kleinman S. G.,  Hall  D. N. B., 1986, ApJS,  62, 501

\bibitem[\protect\citeauthoryear{Kotak et al.}{2005}]{kotak5} Kotak R., Meikle P., van Dyk S. D., 
                 H\"oflich P. A., Mattila S., 2005, ApJ, 628, L123

\bibitem[\protect\citeauthoryear{Kotak et al.}{2009}]{kotak9} Kotak R., et al., 2009, ApJ,704, 306

\bibitem[\protect\citeauthoryear{Kotak et al.}{2006}]{kotak6}  Kotak et al. 2006, ApJ, 651, L117

\bibitem[\protect\citeauthoryear{Lazzati \& Heger}{2016}]{lazzati} Lazzati D., Heger A., 2016, ApJ, 817, 134

\bibitem[\protect\citeauthoryear{Lepp, Dalgarno \& McCray}{Lepp et al.}{1990}]{lepp} Lepp S.,
                Dalgarno A., McCray R.,  1990, ApJ, 358, 262

\bibitem[\protect\citeauthoryear{Liu \& Dalgarno}{1995}]{LD} Liu W., Dalgarno A., 1995, ApJ, 454, 472

\bibitem[\protect\citeauthoryear{Liu, Dalgarno \& Lepp}{Liu et al.}{1992}]{wliu} Liu W., Dalgarno A.,  Lepp S., 1992, ApJ,  396, 679

\bibitem[\protect\citeauthoryear{Liu et al.}{2016}]{nliu} Liu N., Nittler L. R.,  Alexander C. M. O'D., Wang J,
         Pignatari M., Jos\'e J. Nguyen A., 2016, ApJ, 820, 140

\bibitem[\protect\citeauthoryear{Lucy et al.}{1991}]{lucy} Lucy L. B., Danziger I. J., Gouiffes C., Bouchet P., 1991, in
            Supernovae. The Tenth Santa Cruz Workshop in Astronomy and Astrophysics, p.~82
            ed S. E. Woosley, Springer-Verlag, New York

\bibitem[\protect\citeauthoryear{Marion et al.}{2014}]{b21} Marion G. H., et al., 2014, ApJ, 781, 69

\bibitem[\protect\citeauthoryear{Marion et al.}{2015}]{b23} Marion G. H., et al., 2015, ApJ, 798, 39

\bibitem[\protect\citeauthoryear{Marples, Bock \& Parker}{2016}]{b25} Marples P., Bock G., Parker S., 2016, ATel, 8651

\bibitem[\protect\citeauthoryear{Matsuura et al.}{2011}]{matsuura11} Matsuura M., et al., 2011, Science, 333, 1258

\bibitem[\protect\citeauthoryear{Matsuura et al.}{2015}]{matsuura15} Matsuura M., et al., 2015, ApJ, 800, 50

\bibitem[\protect\citeauthoryear{Matsuura et al.}{2017}]{matsuura17} Matsuura M., et al., 2017, MNRAS, 469, 3347

\bibitem[\protect\citeauthoryear{Maurer et al.}{2010}]{maurer} Maurer I., Mazzali P. A., Taubenberger S., Hachinger S.,
        2010, MNRAS, 409, 1441

\bibitem[\protect\citeauthoryear{McGregor \& Hyland}{1987}]{mcgregor} McGregor P. J., Hyland A. R., 1987, IAUC, 4468

\bibitem[\protect\citeauthoryear{Meikle et al.}{1989}]{meikle0} Meikle W. P. S., Spyromilio J., Varani G.-F., Allen D. A., 1989, MNRAS, 238, 193

\bibitem[\protect\citeauthoryear{Meikle et al.}{1993}]{meikle} Meikle W. P. S., Spyromilio J., Allen D. A., 
           Varani G.-F., Cumming R. J., 1993, MNRAS, 261 535

\bibitem[\protect\citeauthoryear{Meikle et al.}{2006}]{meikle06} {Meikle W. P. S., et al., 2006, ApJ, 649, 332}

\bibitem[\protect\citeauthoryear{Meikle et al.}{2007}]{meikle07} {Meikle W. P. S., et al., 2007, ApJ, 665, 608}

\bibitem[\protect\citeauthoryear{Morgan \& Edmunds}{2003}]{morgan} Morgan H. L., Edmunds M. G., 2003, MNRAS, 343, 427

\bibitem[\protect\citeauthoryear{Nittler \& Hoppe}{2005}]{nittler} Nittler L. R.,  Hoppe P., 2005, ApJ, 631, L89

\bibitem[\protect\citeauthoryear{Olivia, Moorwood \& Danziger}{1987}]{oliva} Oliva E., Moorwood A. F., Danziger I. J., 1987, IAUC 4484

\bibitem[\protect\citeauthoryear{Osterbrock \& Ferland}{2006}]{AGN2} Osterbrock D. E., Ferland G. J., 2006, 
Astrophysics of gaseous nebulae and active galactic nuclei, 
Second Edition,  University Science Books, Sausalito, California

\bibitem[\protect\citeauthoryear{Patat et al.}{2015}]{patat} Patat F., et al., 2015, A\&A, 577, A53

\bibitem[\protect\citeauthoryear{Phillips et al.}{2013}]{phillips} Phillips M. M., et al., 2013, ApJ, 779, 38

\bibitem[\protect\citeauthoryear{Pignatari et al.}{2015}]{pignatari} {Pignatari M., et al., 2015, ApJ, 808, L43}

\bibitem[\protect\citeauthoryear{Pozzo et al.}{2006}]{pozzo} Pozzo M., Meikle W. P. S., Rayner J. T., Joseph R. D., Filippenko A. V., 
        Foley R. J., Li W., Mattila S., Sollerman J., 2006, MNRAS, 368. 1169

\bibitem[\protect\citeauthoryear{Raj et al.}{2012}]{raj12} Raj A., Ashok N. M., Banerjee D. P. K., Munari U.,
        Valisa P., Dallaporta S., 2012, MNRAS, 425, 2576
        
\bibitem[\protect\citeauthoryear{Raj et al.}{2015}]{raj15} Raj A., Banerjee D. P. K., Ashok N. M., Kim Sang Chul, 2015, RAA,
15, 993

\bibitem[\protect\citeauthoryear{Rayner et al.}{2003}]{b36} Rayner J. T., Toomey D. W., Onaka P. M., Denault A. J.,
Stahlberger W. E., Vacca W. D., Cushing M. C., Wang, S., 2003, PASP, 115, 362

\bibitem[\protect\citeauthoryear{Rho et al.}{2009}]{rho09} Rho J., Jarrett T. H., Reach W. T., Gomez H., Andersen M., 2009, 693, L39

\bibitem[\protect\citeauthoryear{Rho et al.}{2012}]{rho12} Rho J., Onaka T., Cami J., Reach W. T., 2012,
             ApJ, 747, L6

\bibitem[\protect\citeauthoryear{Rho et al.}{2018a}]{rho18a}  Rho J., Gomez H. L., Boogert A., Smith M. W. L., 
          Lagage, P.-O., Dowell D., Clark C. J. R., Peters E., Cami J., 2018a, MNRAS, 479, 5101

\bibitem[\protect\citeauthoryear{Rho et al.}{2018b}]{rho18} Rho J., Geballe T. R., Banerjee D. P. K., Dessart L., Evans A., Joshi V., 2018b, ApJ, in press
         (arXiv:1808.00683)

\bibitem[\protect\citeauthoryear{Richmond et al.}{1994}]{b27} Richmond M. W., Treffers R, R., Filippenko A, V.,
           Paik Y., Leibundgut B., Schulman E., Cox C. V., 1994, AJ, 107, 1022

\bibitem[\protect\citeauthoryear{Roche, Aitken \& Smith}{Roche et al.}{1993}]{roche} Roche P. F., Aitken D. A., Smith C. H., 1993, MNRAS, 261, 522

\bibitem[\protect\citeauthoryear{Sand et al.}{2016}]{b29} Sand D. J., et al., 2016, ApJ, 822, L16

\bibitem[\protect\citeauthoryear{Sarangi \& Cherchneff}{2013}]{sarangi2} {Sarangi A., Cherchneff I., 2013, ApJ, 776, 107}

\bibitem[\protect\citeauthoryear{Sarangi \& Cherchneff}{2015}]{sarangi3} Sarangi A., Cherchneff I., 2015, A\&A, 575, A95

\bibitem[\protect\citeauthoryear{Sarangi, Matsuura \& Micelotta}{Sarangi et al.}{2018}]{sarangi}
        Sarangi A., Matsuura M., Micelotta E. R., 2018, Sp. Sci. Rev., 214, 63

\bibitem[\protect\citeauthoryear{Schlafly \& Finkbeiner}{2011}]{b31} Schlafly E., Finkbeiner D. P.,
2011, ApJ, 737, 103

\bibitem[\protect\citeauthoryear{Sembach, Danks \& Savage}{1993}]{b33} Sembach K. R., Danks A. C.,
Savage B. D., 1993, A\&A, 100, 107

\bibitem[\protect\citeauthoryear{Sharp \& Hoeflich}{1989}]{sharp2} Sharp C., H\"oflich P., 1989, Highlights of Astronomy, 8, 207

\bibitem[\protect\citeauthoryear{Sharp \& Hoeflich}{1990}]{sharp} Sharp C., H\"oflich P., 1990, Ap\&SS, 171, 213 

\bibitem[\protect\citeauthoryear{Smith et al.}{2011}]{smith} Smith N., et al., 2011, MNRAS, 418, 1959

\bibitem[\protect\citeauthoryear{Spyromilio et al.}{1988}]{b40} Spyromilio J., Meikle W. P. S., Learner R. C. M., Allen D. A., 1988, Nature, 334, 327

\bibitem[\protect\citeauthoryear{Spyromilio \& Leibundgut}{1996}]{SL} Spyromilio J., Leibundgut B., 1996, MNRAS, 283, L89

\bibitem[\protect\citeauthoryear{Spyromilio, Leibundgut \& Gilmozzi}{Spyromilio et al.}{2001}]{b42} 
            Spyromilio J., Leibundgut B., Gilmozzi R., 2001, A\&A, 376, 188

\bibitem[\protect\citeauthoryear{Srivastava et al.}{2015}]{b35} Srivastava M., Ashok N. M.,
           Banerjee D. P. K.,  Sand D., 2015, MNRAS, 454, 1297

\bibitem[\protect\citeauthoryear{Srivastava, Banerjee \& Ashok}{Srivastava et al.}{2016}]{b37} Srivastava M., Banerjee
          D. P. K., Ashok N. M., 2016, ATel 8686 

\bibitem[\protect\citeauthoryear{Starrfield, Gehrz \& Truran}{1997}]{starrfield} Starrfield S., Gehrz R. D., 
          Truran J. W., 1997, in Bernatowicz T. J., Zinner E., eds, AIP Conf. Proc. 402, 
          Astrophysical Implications of the Laboratory Study of Presolar Materials, American Institute of Physics, New York

\bibitem[\protect\citeauthoryear{Strittmatter et al.}{1977}]{stritt} {Strittmatter P. A., Woolf N. J., 
         Thompson R. I., Wilkerson S., 
         Angel J. R. P., Stockman H. S., Gilbert G., Grandi S. A., Larson H., Fink U., 1977, ApJ, 216, 23}
          
\bibitem[\protect\citeauthoryear{Stritzinger et al.}{2016}]{b39} Stritzinger M., et al. 2016, ATel 8657

\bibitem[\protect\citeauthoryear{Strutskie et al.}{2006}]{2mass} Strutskie M. F., et al., 2006, AJ, 131, 1163

\bibitem[\protect\citeauthoryear{Sugarman \& Lawrence}{2016a}]{b43a} Sugarman B., Lawrence S., 2016a, ATel 8759

\bibitem[\protect\citeauthoryear{Sugarman \& Lawrence}{2016b}]{b44b} Sugarman B., Lawrence S., 2016b, ATel 8890

\bibitem[\protect\citeauthoryear{Taubenberger et al.}{2011}]{b45} Taubenberger S., et al. 2011, MNRAS, 413, 2140

\bibitem[\protect\citeauthoryear{Temim et al.}{2017}]{temim} Temim T., Dwek E., Arendt R. G., Borkowski K. J., Reynolds S. P., 
              Slane P., Gelfand J. D., Raymond J. C., 2017, ApJ, 836, 129

\bibitem[\protect\citeauthoryear{Thomas et al.}{2016}]{b44} Thomas A., et al. 2016, ATel 8664

\bibitem[\protect\citeauthoryear{Todini \& Ferrara}{2001}]{todini} Todini P., Ferrara A., 2001, MNRAS, 325, 726

\bibitem[\protect\citeauthoryear{Vacca, Cushing \& Rayner}{2003}]{vacca} Vacca W. D., Cushing M. C., Rayner J. T., 2003, PASP, 115, 389

\bibitem[\protect\citeauthoryear{Valenti et al.}{2008}]{b49} Valenti S.,  et al., 2008, ApJ, 673, L155.

\bibitem[\protect\citeauthoryear{Van Dyk et al.}{2016}]{vandyk} Van Dyk S. D., et al., 2016, ATel 8693

\bibitem[\protect\citeauthoryear{Yi et al.}{2016}]{b47} Yi Weimin, Zang Ju-Jia,
Wu Xue-Bing, Shappee, B. J., Prieto, J. L., Subo D., 2016, ATel 8655

\bibitem[\protect\citeauthoryear{Williams}{1992}]{williams} Williams R. E., 1992, AJ, 104, 725

\bibitem[\protect\citeauthoryear{Wooden et al.}{1993}]{wooden} Wooden D. H., Rank D. M., Bregman J, D., Witteborn F. C.. Tielens
            A. G. G. M., Cohen M., Pinto P. A., Axelrod T. S., 1993, ApJS, 88, 477

\bibitem[\protect\citeauthoryear{Yang et al.}{2016}]{yang} Yang Y., et al., 2016, ATel 8668


\end{thebibliography}
\end{document}